\documentclass[11pt]{article}
\usepackage{amsfonts,amsmath}
\usepackage{psfrag}
\usepackage{graphics}
\usepackage{subfigure}
\usepackage{ifpdf}
\usepackage{epsf}
\ifpdf
\usepackage{epstopdf}
\usepackage{hyperref}
\else
\usepackage[hypertex]{hyperref}
\fi

\newcommand{\scc}{c_{\psi\psi\psi}}

\advance\voffset by -1.6cm
\advance\hoffset by -1.7cm
\newcommand\sect[1]{\section{#1}\setcounter{equation}0} 

\newcommand\be            {\begin{equation}}
\newcommand\bea           {\begin{equation}\begin{array}l\displaystyle}
\newcommand\ee            {\end{equation}}
\newcommand\eear          {\end{array}}
\newcommand\enl           {\\[1em]\displaystyle}
\newcommand\etb           {& \displaystyle}
\newcommand\erf[1]        {\eqref{#1}}
\newcommand\labl[1]       {\label{#1}\ee}

\newcommand\eps           {\varepsilon}
\newcommand\fii           {\varphi}

\newcommand\id            {{\rm id}}
\newcommand\ol            {\overline}

\newcommand\rrangle       {\rangle\!\rangle}
\newcommand\Vir           {\text{Vir}}

\newcommand\Zb            {\mathbb{Z}}

\newcommand\Hc            {\mathcal{H}}
\newcommand\Ic            {\mathcal{I}}

\newcommand\updown[2]{{#1}_{#2}}
\newcommand\kapl          {\kappa_l}
\newcommand\kapr          {\kappa_r}
\newcommand\laml          {\lambda_l}
\newcommand\lamr          {\lambda_r}
\newcommand\lamb[1]{\updown{\lambda}{#1}}
\newcommand\kapp[1]{\updown{\kappa}{#1}}
\newcommand\kappp[2]{\kappa_{#1}^{#2}}
\newcommand\mul          {\mu_l}
\newcommand\mur          {\mu_r}
\newcommand\bt[1]{{\updown{\tilde\beta}{#1}}}
\newcommand\bb[1]{{\updown{\beta}{#1}}}
\newcommand\gamm[1]{{\updown{\gamma}{#1}}}

\renewcommand\vec[1]{{\vert{#1}\rangle}}
\newcommand\vecc[1]{{\vert\hspace{1pt}#1\hspace{1pt}\rrangle}}
\newcommand\vvecc[1]{{\vert\hspace{-1pt}\vert\hspace{1pt}#1\hspace{1pt}\rrangle}}
\newcommand\cev[1]{{\langle{#1}\vert}}
\newcommand\ccevv[1]{{\langle\!\langle\hspace{1pt}#1\hspace{1pt}\vert\hspace{-1pt}\vert}}

\def\3pt#1#2#3{{\langle{#1}|{#2}|{#3}\rangle}}

\def\thefootnote{\fnsymbol{footnote}}
\begin{document}

\begin{flushright}  {~} \\[-12mm]
{\sf KCL-MTH-09-06}\\[1mm]
{\sf 0907.1497 [hep-th]}
\end{flushright} 

\thispagestyle{empty}

\begin{center} \vskip 14mm
{\Large\bf Defect flows in minimal models}\\[20mm] 
{\large 
M\'arton Kormos$^{a,b,c}$,\,  
Ingo Runkel$^{a}$,\, 
G\'erard M.\ T.\ Watts$^{a}$~~\footnote{Emails: 
{\tt kormos@sissa.it, Ingo.Runkel@kcl.ac.uk, Gerard.Watts@kcl.ac.uk}}}
\\[8mm]
\it$^a$ Dept.\ of Mathematics, King's College London,\\
Strand, London WC2R\;2LS, UK
\\[8mm]
$^b$ International School for Advanced Studies (SISSA),\\
Via Beirut 4, 34014 Trieste, Italy
\\[8mm]
$^c$ Istituto Nazionale di Fisica Nucleare, Sezione di Trieste, Italy

\vskip 22mm
\end{center}

\begin{quote}{\bf Abstract}\\[1mm]
In this paper we study a simple example of a two-parameter space of
renormalisation group flows of defects in Virasoro minimal models. We
use a combination of exact results, perturbation theory and the
truncated conformal space approach to search for fixed points and
investigate their nature. For the Ising model, we confirm the recent
results of Fendley et al. In the case of central charge close to one, we
find six fixed points, five of which we can identify in terms of known
defects and one of which we conjecture is a new non-trivial conformal
defect. We also include several new results on exact properties of
perturbed defects and on the renormalisation group in the truncated
conformal space approach.
\end{quote}

\vfill
\newpage 

\setcounter{footnote}{0}
\def\thefootnote{\arabic{footnote}}

\tableofcontents

\sect{Introduction}

By a defect in a two-dimensional conformal field theory we mean a line
of inhomogeneity on the surface, where the expectation values of
fields are allowed to be discontinuous or even singular. 
An example would be the continuum limit of a lattice model where the
couplings are altered from their normal values along a line.  

A typical defect is not invariant under a scale transformation and
this leads to an action of the renormalisation group on the space of
defects. The fixed points of the renormalisation group are clearly of
interest and these are the conformal defects. 
The problem of studying conformal defects is equivalent to
studying general conformal boundary conditions of a folded model
\cite{Wong:1994pa}. Even if one starts from a rational conformal field
theory, the  
folded model will typically no longer be rational with respect to the
diagonal symmetry preserved by the boundary condition corresponding to
the defect. This means that the representation theoretic methods used
to construct the bulk theory cannot be applied to get a handle on the
defect itself.  
There are, however, two
distinguished subsets of conformal defects which are known as
topological (or purely transmitting) defects and factorising (or purely
reflecting defects), which are much easier to study than the general
case.

Defects have an obvious generalisation to interfaces between different
conformal field theories, possibly of different central charge.  A
particular model may, in fact, be simple enough to allow one to
classify all conformal defects or interfaces; this is the case for
defects in the Lee-Yang model \cite{Quella:2006de} and in the critical
Ising model \cite{Oshikawa:1996dj}, and for interfaces between the
Lee-Yang and the Ising model \cite{Quella:2006de}.  Particular defects
may preserve a rational sub-algebra in the folded model
\cite{Quella:2002ct,Quella:2006de,Fredenhagen:2006qw}, or a
semi-classical analysis may suggest the existence of distinguished
conformal defects and interfaces, for example in a WZW model for a
given group \cite{Bachas:2004sy} or between such WZW models at
different levels \cite{Fredenhagen:2005an}.

One may also think of interfaces as `symmetries' which relate the
properties of the two theories they link. For example, interfaces
provide group symmetries and order-disorder dualities
\cite{Frohlich:2004ef}, they relate different renormalisation group
flows of boundary conditions \cite{Graham:2003nc}, there is a
preferred interface joining the UV and IR fixed point of a given
quantum field theory \cite{Brunner:2007ur}, and they can be used as
spectrum generating symmetries in string theory \cite{Bachas:2008jd}.
Interfaces can also be related to tunnelling in the quantum Hall
effect \cite{Fendley:2009gm}.  Defects (and also interfaces) can be
composed by placing the defect lines parallel to each other and
letting their distance tend to zero
\cite{Petkova:2000ip,Frohlich:2006ch,Brunner:2007qu,Runkel:2007wd,Bachas:2007td}.
This process may or may not be singular, but if it can be defined then it
describes an algebraic structure on the space of conformal field
theories which deserves more investigation.

\begin{figure}[t]

$$  
  \begin{picture}(160,160)
  \put(0,0){\scalebox{0.80}{\includegraphics{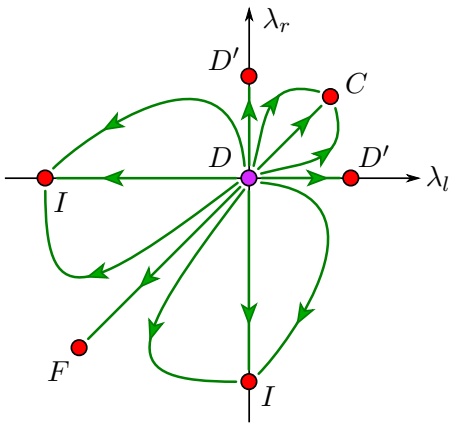}}}
  \put(0,0){
     \setlength{\unitlength}{.80pt}\put(-90,-587){
     \put(290,700)   {$\laml$}
     \put(213,775)   {$\lamr$}
     \put(252,744)   {$C$}
     \put(111,608)   {$F$}
     \put(187,709)   {$D$}
     \put(187,755)   {$D'$}
     \put(258,709)   {$D'$}
     \put(212,596)   {$I$}
     \put(114,688)   {$I$}
     }\setlength{\unitlength}{1pt}}
  \end{picture}
$$

\caption{\small The proposed flows for the perturbation \eqref{eq:MM-12-pert}
  for $p>3$. The point $D$ is the $(1,2)$-defect. The
  possible endpoints are: $I$ -- the identity defect, $D'$ --
  the $(2,1)$-defect, $F$ -- a factorising defect given by the sum
  $\sum_{r=1}^{p-1} \vvecc{r,1} \ccevv{r,1}$ of $p{-}1$ conformal
  boundary conditions, and finally $C$ -- the new conformal
  defect. For details see the body of the paper.} 
\label{fig:flow-p>3}
\end{figure}

In this paper we study a very simple example of a two-parameter space
of renormalisation group flows.  We identify the fixed points of these
flows with the aim of finding new non-trivial conformal defects.  The
starting point for our flows is a particular topological defect $D$ in
the Virasoro minimal model $M(p,p+1)$ and the two parameters of our
space are $\laml$ and $\lamr$, describing perturbations of $D$.  We
arrive at the following conjecture for the space of flows for $p>3$ in
the neighbourhood of this defect, shown in figure \ref{fig:flow-p>3}.

Note that, in this figure, $\laml$ and $\lamr$ represent renormalised coupling
constants. An example of such coupling constants are the coupling
constants used in the TCSA scheme in section \ref{sec:TCSA}
taken at some fixed finite cut-off.

Four
of the fixed points (labelled $D'$ and $I$) can be identified exactly 
and are topological defects; one can be studied numerically and we
conjecture that it is a factorising defect $F$; finally we
conjecture the existence of a new non-trivial conformal defect $C$
which we study both numerically and perturbatively. 

A renormalisation group analysis in the case of WZW models showed a
similar non-trivial fixed point for $\laml=\lamr\neq0$, which is a
good candidate for a non-topological conformal defect
\cite{Bachas:2004sy}.

The case $p=3$ has recently been studied in \cite{Fendley:2009gm} and solved
exactly by relating it to a free fermionic model. The space
of RG flows is qualitatively different but the fixed
points corresponding to $I, C, D'$ and $F$ are present in this model
and agree with our conjectures for the exact forms of $I$, $F$ and
$D'$. Our numerical calculations agree with the results of
\cite{Fendley:2009gm}, confirming the validity of our numerical method.

Unfortunately, we have so far been unable to calculate
exactly or numerically other characteristic quantities of the new
conformal defect $C$, such as its $g$-value \cite{Affleck:1991tk} (defined
to be that of the corresponding conformal boundary in the folded
model) or its reflection coefficient \cite{Quella:2006de}. However, we
can calculate the $g$-value perturbatively for large values of $p$.

\medskip

This paper is organised as follows. In section \ref{sec:topdef} we
collect the properties of topological defects needed in the subsequent
analysis.  Section \ref{sec:exact} contains several exact results on
defects, some new to this paper.  Section \ref{sec:pert} contains the
renormalisation group analysis of the perturbation for large $p$ and
identifies the new conformal defect as a perturbative fixed point. The
numerial truncated conformal space approach used 
to support the proposed flows of figure \ref{fig:flow-p>3}
is described in section 
\ref{sec:TCSA} with the results given in section
\ref{sec:TCSA-results}. 
Finally, section
\ref{sec:sum} contains our conclusions.

\sect{Topological, factorising and conformal defects}\label{sec:topdef}

To derive the properties of conformal defects, 
consider the
complex plane with a defect line placed on the real axis. The defect
preserves the conformal symmetry of the bulk theory if the field
$T_{xy} = \tfrac{i}{2\pi}(T-\bar T)$ is continuous across the real
axis. 

This condition is unchanged if we now consider the defect on a
cylinder obtained by identifying $z$ with $z+2\pi$. One can map this
cylinder to the whole plane by $z \mapsto \exp(i z)$ and the real line
gets mapped to the unit circle. In this formulation, a defect on the
unit circle defines an operator $D$ on the space of states $\Hc$ in
radial quantisation. Translating the condition to this situation, a
defect is conformal if the operator $D$ commutes with the difference
of the holomorphic and anti-holomorphic copy of the Virasoro modes,
\be
  [L_m - \bar L_{-m} , D ] = 0 \quad \text{for~all}~m \in \Zb~.
\labl{eq:LL-D-conf}

General conformal defects of a given conformal field theory are
difficult to describe because according to \eqref{eq:LL-D-conf} they
only preserve the diagonal Virasoro algebra $L^d_m = L_m - \bar
L_{-m}$ (with central charge $c^d = c+\bar c$) of the full holomorphic
and anti-holomorphic symmetry. 

\begin{figure}[t]

$$  
  \begin{picture}(450,10)
  \put(0,0){\scalebox{0.32}{\includegraphics{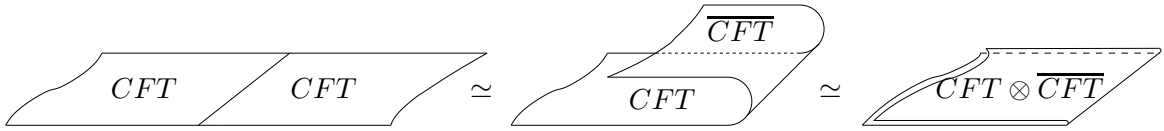}}}
  \put(0,0){
     \setlength{\unitlength}{.80pt}\put(-90,-587){
     \put(140,600)   {$CFT$}
     \put(225,600)   {$CFT$}
     \put(310,600)   {$\simeq$}
     \put(385,595)   {$CFT$}
     \put(422,627)   {$\overline{CFT}$}
     \put(475,600)   {$\simeq$}
     \put(530,600)   {$CFT\otimes\overline{CFT}$}
     }\setlength{\unitlength}{1pt}}
  \end{picture}
$$

\caption{The equivalence between a defect in a CFT and a boundary
  condition in the folded model}
\label{fig:fold1}
\end{figure}

A very useful way to think of this condition is in terms boundary
conditions on the folded model. If one considers a CFT on the complex
plane and folds the worldsheet over along the imaginary axis, the
resulting model consists of the tensor product $CFT \otimes
\overline{CFT}$ on the half plane with a boundary condition inserted
along the imaginary axis as shown in figure \ref{fig:fold1}, where
$\overline{CFT}$ is the original CFT with holomorphic and
anti-holomorphic dependences swapped.
If the original model had a conformal defect
along this line, 
\eqref{eq:LL-D-conf} implies that it is a conformally invariant
boundary condition in the folded model. Note that this includes the
possibility that there 
is no defect (or the identity defect) so that there is a distinguished
conformal boundary condition corresponding to the absence of any
defect. This correspondence between defects in a CFT and boundary
conditions on the folded model $CFT \otimes \overline{CFT}$ will often
be used in what follows.

As said already, even if one starts
from a rational model, the representation theoretic methods used
to construct the bulk theory cannot be applied to get a handle on a
general conformal defect. There are, however, two distinguished
subsets of conformal defects which are much easier to study than the
general case and which we consider now.

\subsection{Topological defects and defect fields}
\label{ss:topdef}

\medskip

Topological defects are a particular class of conformal defects which
preserve a larger symmetry than the diagonal Virasoro symmetry and 
are amenable to classification on account of this.
The larger symmetry in question is, in fact, the full bulk symmetry,
that is, the conformal defect obeys the stronger condition 
\be
  [L_m,D] = 0 = [\bar L_m,D] \quad \text{for~all}~m \in \Zb~,
\labl{eq:LL-D-top}
which clearly implies \eqref{eq:LL-D-conf}. Such defects are called
topological because they are tensionless and can be deformed on the
surface without affecting the value of correlators. Topological
defects were first studied in \cite{Petkova:2000ip} (and the name
`topological' was introduced in \cite{Bachas:2004sy}). Condition
\eqref{eq:LL-D-top} is equivalent to demanding that $T$ and $\bar T$
are separately continuous across the defect line.  As a consequence,
the space of defect fields forms a representation of the holomorphic
and anti-holomorphic copy of the Virasoro algebra, just as the space
of bulk fields. In particular, a defect field $\phi$ has a left and
right conformal weight $(h_l,h_r)$.  

The classification of topological defects can be carried out for
rational conformal field theories and general modular invariants if
one requires the defects to preserve the holomorphic and
anti-holomorphic copy of the chiral algebra
\cite{Petkova:2000ip,Fuchs:2002cm}. Here we consider only the diagonal
unitary Virasoro minimal models, for which the exposition simplifies.  

Let thus $M(p,p{+}1)$ be the Virasoro minimal model of central charge
$c=1-6/(p^2+p)$ and with diagonal modular invariant partition
function.  The irreducible representations $R_i$ of the Virasoro
algebra $\Vir$ occurring in these models are labelled by the set of
Kac labels $i \in \Ic_p$,
\be
  \Ic_p = \big\{ (r,s) \big| 1 \le r \le p{-}1, 1 \le s \le p \big\}
  / \sim
  \quad \text{where} 
  ~ (r,s) \sim (p{-}r,p{+}1{-}s)~.
\ee
The space of states propagating on a cylinder, or equivalently the
space of bulk fields, decomposes into representations of $\Vir \oplus
\ol\Vir$ as 
\be
  \Hc = \bigoplus_{i \in \Ic_p} R_i \otimes \bar R_i~.
\ee
Both the elementary conformal boundary conditions \cite{Cardy:1989ir}
and the elementary topological defects \cite{Petkova:2000ip} in
$M(p,p{+}1)$ are labelled by the set $\Ic_p$. Let $\vvecc{a}$ be the
boundary state corresponding to removing the open unit disc from the
complex plane and labelling the resulting boundary by $a \in
\Ic_p$. Denote by $D_k : \Hc \rightarrow \Hc$ the operator describing
a topological defect with label $k \in \Ic_p$ placed on the unit
circle in the complex plane. Explicitly, these two quantities are
given by 
\be
  \vvecc{a} = \sum_{i \in \Ic_p} 
  \frac{S_{ai}}{\sqrt{S_{0i}}} \, \vecc{i}
  \quad \text{and} \quad
  D_k = \sum_{i \in \Ic_p} \, \frac{S_{ki}}{S_{0i}}  \,
  id_{R_i \otimes \bar R_i}~,
\labl{eq:bndstate-defectop}
where $0$ is the identity or vacuum representation $(1,1)$, $\vecc{i}$
is the Ishibashi state in the (algebraic completion of) $R_i \otimes
\bar R_i$ and $S_{ij}$ is the modular $S$-matrix for $M(p,p{+}1)$, 
\be
  S_{(r,s)\,(x,y)} = \sqrt{8/(p^2{+}p)} \cdot
  (-1)^{1+sx+ry} \cdot
  \sin \frac{\pi (p{+}1) rx}{p} \cdot
  \sin \frac{\pi p sy}{p{+}1}~.
\labl{eq:min-mod-S}
An important property of a defect is its entropy $g$ defined in the same
way as the boundary entropy and identical in value to the boundary
entropy of the corresponding boundary condition in the folded model
\cite{Affleck:1991tk}. 
From \eqref{eq:bndstate-defectop} we read off
the $g$-value of a boundary condition and a topological defect as
\be
g(\,\vvecc{a}) = \frac{S_{a0}}{\sqrt{S_{00}}}
  \quad \text{and} \quad
g(D_k) = \frac{S_{k0}}{S_{00}}
\;.
\labl{eq:gvalues-fact}

We recall that topological defects can be deformed freely on the surface
as long as they do not cross field insertions, boundaries, or defect
lines. The fusion of two topological defects corresponds to the 
composition of the defect operators, and the fusion of a defect with a 
boundary condition is given by the action of the defect operator on
the boundary state. One easily checks that for $i,j,k,a \in \Ic_p$,
\be
  D_{(1,1)} = id^{\phantom{o}}_{\Hc}
  ~~,~~~~
  D_k \, D_l = \sum_{m \in \Ic_p} N_{kl}^{~m} \, D_m
  ~~,~~~~
  D_k \, \vvecc{a} = \sum_{b \in \Ic_p} N_{ka}^{~b} \, \vvecc{b}~,
\ee
where $N_{ij}^{~k}$ are the fusion rule coefficients.
By evaluating the corresponding partition functions, or by using the
methods of \cite{Frohlich:2006ch}, one finds that the space of defect
fields living on a defect labelled by $k \in \Ic_p$ is given by 
\be
  \Hc_k^D = \bigoplus_{i,j \in \Ic_p} \big(R_i \otimes \bar R_j\big)^{\oplus \sum_{x \in \Ic_p} N_{ij}^{~x} N_{kk}^{~x}}~.
\labl{eq:defect-fields}
We will also need the space of states propagating on a strip with
boundary condition $(1,1)$ on one side and $a \in \Ic_p$ on the other
side, and with a defect labelled $k\in\Ic_p$ running parallel to the
boundaries inside the strip. In the same way as
\eqref{eq:defect-fields} one finds this is 
\be
  \Hc^{(1,1),a}_k
  = \bigoplus_{i \in \Ic_p} R_i^{\oplus N_{ak}^{~i}} ~.
\labl{eq:strip-states}

\subsection{Factorising defects}

\begin{figure}[t]

$$  
  \begin{picture}(300,10)
  \put(0,0){\scalebox{0.32}{\includegraphics{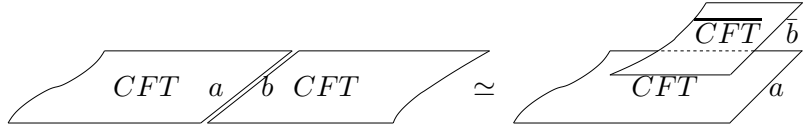}}}
  \put(0,0){
     \setlength{\unitlength}{.80pt}\put(-90,-587){
     \put(140,600)   {$CFT$}
     \put(185,600) {$a$}
     \put(210,600) {$b$}
     \put(225,600)   {$CFT$}
     \put(310,600)   {$\simeq$}
     \put(385,600)   {$CFT$}
     \put(415,625)   {$\overline{CFT}$}
     \put(450,600) {$a$}
     \put(458,625) {$\bar b$}
     }\setlength{\unitlength}{1pt}}
  \end{picture}
$$

\caption{The equivalence between a factorising defect in a CFT and
  separate boundary conditions in the folded model}
\label{fig:fold-fact}
\end{figure}

By a factorising defect we mean a conformal defect represented by an
operator $F$ that 
satisfies the stronger conditions 
\be
  (L_m - \bar L_{-m}) F = 0 = F (L_m - \bar L_{-m})
  \quad \text{for~all}~m \in \Zb~,
\ee
which then imply \eqref{eq:LL-D-conf}. 
Such a defect is  totally reflecting
and is simply a
sum of products of conformal boundary states 
\be
  F = \sum_{a,b \in \Ic_p} n_{ab} \,\vvecc{a}\,\ccevv{b}
\labl{eq:Fform}
for some non-negative integers $n_{ab}$. In the folded model this
corresponds to separate boundary conditions on the two sheets, as in
figure \ref{fig:fold-fact}.
The $g$-value of the factorising defect \eqref{eq:Fform} is 
\be
g(F) = \sum_{a,b\in \Ic_p} n_{ab} \frac{S_{a0}S_{b0}}{S_{00}}
\;.
\labl{eq:gvalues-top}

\subsection{Conformal defects}\label{sec:confdefs}

Returning to general conformal defects, we give the form of the
partition function on a cylinder and define a useful
pairing on such defects.

Let $d$ and $d'$ label two conformal defects and consider a torus of
width $R$ and height $L$ with two vertical defect lines at distance
$\eta R$, $0<\eta<1$, from each other. Denote by $D_d$ and $D_{d'}$
the defect operators of the defects $d$ and $d'$. Define the partition
function 
\be
  Z_{d,d'}(q;\eta) = \text{tr}_{\Hc}\Big(  \,
  {\tilde q}^{\,\eta(L_0+\bar L_0 - c/12)} \, (D_d)^\dagger \, 
  {\tilde q}^{(1-\eta)(L_0+\bar L_0 - c/12)} \, D_{d'} \, \Big)
\labl{eq:Zdd-def}
where $q=e^{- 2 \pi L / R}$ and $\tilde q = e^{- 2 \pi R / L}$. The
adjoint $(D_d)^\dagger$ describes the defect operator for the defect
$d$ inserted with reversed orientation (see \cite{Petkova:2000ip} and
\cite[sect.\,6.1]{Fuchs:2007tx}). 
Using the explicit expressions \eqref{eq:bndstate-defectop} one checks
that for the factorising and topological defects in $M(p,p{+}1)$ one
has 
\be
  (\vvecc{a}\ccevv{b})^\dagger = \vvecc{b}\ccevv{a}
  \quad \text{and} \quad
  (D_k)^\dagger = D_k ~.
\ee

In the special case that $\eta = \tfrac12$, we can identify the
original CFT on a torus with two defects with the
folded model on a cylinder of circumference $L$ and width $R/2$ with
conformal boundary conditions corresponding to the defects $d$ and
$d'$ at the ends of the cylinder.
The partition function of the folded model on a cylinder
of circumference $L$ and width $R/2$ is a sum of Virasoro characters
$\chi_{h,2c}(q)$ at twice the central charge of $M(p,p{+}1)$, so that
the partition function can be written as
\be
  Z_{d,d'}(q;\tfrac12) = 
  \sum_h m_h \, \chi_{h,2c}(q) 
\labl{eq:Zfolded}
where $q=e^{- 2 \pi L / R}$, $m_h$ is the multiplicity of the
representation with highest weight $h$ and the sum can be over a
finite or infinite set of weights $h$.
We have assumed that the spectrum of the folded model on the strip is
discrete, i.e.\ that $Z_{d,d'}(q;\tfrac12)$ is a sum of powers of $q$,
rather than an integral. This may not be the case for a general
conformal defect, but it is true for the topological defects $D_k$ and
for the totally factorising defects $\vvecc{a}\ccevv{b}$ as can be
checked with the explicit expression in \erf{eq:bndstate-defectop}. We
investigate defects which can be reached by renormalisation group
flows starting at these defects, and so we expect the spectrum at the
IR fixed point to be discrete as well.

We define the pairing $(d,d')$ by
\be
  ( d , d') = m_0 \in \Zb_{\ge 0} ~,
\labl{eq:conf-pair}
i.e.\ the multiplicity of the vacuum character in
$Z_{d,d'}(q;\tfrac12)$.
    For unitary models conformal weights have to be non-negative so that 
    we can write $(d,d')$ as the limit
\be
  ( d , d') = 
  \lim_{q \rightarrow 0} q^{c/12} \, Z_{d,d'}(q;\tfrac12) ~.
\labl{eq:dd-def-limit}
Since $\text{tr}(X) =
(\text{tr}(X^\dagger))^*$ 
for any operator $X$,
and $(d,d')$ is real, 
          it follows from \eqref{eq:dd-def-limit} that 
$(d,d') = (d',d)$. 

By passing to the folded model one checks that $(d,d) \ge 1$, that
$(d,d) = 1$ if and only if the conformal defect $d$ is elementary, and that
$(d,d')=0$ if $d$ and $d'$ are elementary and distinct. The latter statement
follows in the folded model since one cannot have a boundary changing
field of weight zero between two distinct elementary boundary
conditions $a$ and 
$b$ (such a field could be pushed along the $a$ boundary changing it
into a $b$ boundary without affecting correlators, showing that $a$
and $b$ have equal boundary states\footnote{ 
  Some care has to be taken since in general the boundary state does
  not characterise the boundary condition uniquely. Similarly, the
  defect operator does in general not determine the defect
  uniquely. An example of this is discussed in
  \cite{Fuchs:2007tx}. For unitary models this ambiguity is not
  expected to occur, i.e.\ there is a one-to-one correspondence
  between boundary conditions and boundary states, and similar for
  defects.} 
). 

In particular, if $d$ is a superposition and $d'$ is elementary, then
$(d,d')$ gives the multiplicity of $d'$ in the decomposition of $d$
into elementary defects.  

Another useful property of the pairing is that if $k \in \Ic_p$ is a
topological defect, and $k \star d$ denotes the fused defect (which is
again conformal and has the defect operator $D_k D_d$) we have 
\be
  (k \star d,d') = (d,k \star d') 
  \quad \text{and} \quad
  (d \star k,d') = (d,d' \star k) ~.
\labl{eq:move-k-in-pair}
The first equality follows from 
\bea
\text{tr}_{\Hc}\Big(  \,
  {\tilde q}^{\frac12(L_0+\bar L_0 - c/12)} \, (D_k D_d)^\dagger \, 
  {\tilde q}^{\frac12(L_0+\bar L_0 - c/12)} \, D_{d'} \, \Big)
\enl \qquad
=
\text{tr}_{\Hc}\Big(  \,
  {\tilde q}^{\frac12(L_0+\bar L_0 - c/12)} \, (D_d)^\dagger \, 
  {\tilde q}^{\frac12(L_0+\bar L_0 - c/12)} \, D_k D_{d'} \, \Big)
\eear\ee
as $D_k$ is self-adjoint and commutes with $L_0+\bar L_0$.
    The second equality can be seen similarly.

\subsection{Defect perturbations}
\label{sec:chirdefpert}

Finally we consider perturbations of topological defects.
A perturbation of a topological defect by a defect field
$\phi_{h_l,h_r}$ is relevant if $h_l+h_r<1$. This is in contrast to
bulk perturbations, which are integrated over the whole surface, not
just over the defect line, and are relevant if $h_l+h_r<2$. 

The first thing to note is that all topological defects in unitary
minimal models have relevant
perturbations, even the identity defect (that is, no defect), which is
quite different from the boundary situation where there are $(p-1)$
stable boundary conditions for $M(p,p+1)$.
The fields on the identity defect are all scalars and are exactly the
bulk fields $\phi_{h,h}$. There are $\lfloor\sqrt{2p^2+2p+1}\rfloor-2$ of these
with scale dimension less
than or equal to 1, as opposed to $2p-3$ which have scale dimension
$\leq 2$.
In any case, even the identity defect is unstable to defect perturbations by
these bulk fields (considered as fields on the defect).

As another example, on the $(1,2)$ defect there are both scalar primary fields
$\phi_{h,h}$ and non-scalar primary fields $\phi_{h,h'}$ with $h\neq
h'$.  
There are $(p-1)^2$ scalar primary fields of which
$2\lfloor \sqrt{2p^2+2p+1} \rfloor -6$ have dimension $\leq 1$  and
$(p-1)(p-2)$ non-scalar primary fields of which 
$2(p-2)$ have dimension $\leq 1$.  
The relevant non-scalar fields are 
\be
\{ 
 \phi_{(r,r),(r,r+2)} \;\hbox{ and }\;
 \phi_{(r,r+2),(r,r)}
 \;|\; 1\leq r\leq p-2
\}
\;.
\ee

We note here that the interpretation of a defect as a boundary
condition in a folded model means that the boundary $g$-theorem
\cite{Affleck:1991tk,Friedan:2003} applies to perturbed defects as
well: $g$ is  decreasing along RG flows and so restricts the possible
IR fixed points accessible from any given UV fixed point.

The simplest class of relevant perturbations are those by chiral defect
fields $\phi_{h,0}$ or $\phi_{0,h}$ with $h<1$. Such perturbations
were investigated in \cite{Konik:1997,Runkel:2007wd}. These perturbed defects are
believed to have
particularly nice properties, for example to commute
with $L_0+\bar L_0$, 
             see section \ref{sss:trans} below.

On the other hand, since a topological defect
perturbed by $\laml \cdot \phi_{h,0}$, say, will commute with all
$\bar L_m$ for arbitrary values of the coupling $\laml$, the conformal
defect obtained as the IR fixed point for large $\laml$ will
necessarily be topological \cite{Bachas:2004sy}. To obtain an IR fixed
point which is conformal but not topological, the next simplest
perturbation to try is the defect field 
\be
  \laml \cdot \phi_{h,0} ~+~ \lamr \cdot \phi_{0,h} ~.
\labl{eq:defect-pert-field}
A topological defect perturbed by this field will no longer commute
with either the $L_m$'s or the $\bar L_m$'s and in general it will
also not commute with $L_0+\bar L_0$ (there are exceptions, for
example the fusion of a defect perturbed by $\phi_{h,0}$ with a defect
perturbed by  $\phi_{0,h}$).

For the conformal weight $h$ in
\eqref{eq:defect-pert-field} we choose $h_{(1,3)} = (p-1)/(p+1)$. This
field is of interest for several related reasons:
in many circumstances it leads to integrable perturbations which can
be solved exactly
and 
we expect the
purely chiral perturbations to be integrable and exactly solvable;
this perturbation is also related to one of the integrable lattice
description of minimal models; 
         finally, it is
particularly suitable for a
renormalisation group analysis, as 
it does not generate further relevant fields under fusion (note that
$\phi_{h,h}$ already has weight greater than  or equal to 1), and it becomes
marginal in the limit $p\rightarrow\infty$. 

The simplest topological
defect which allows for a perturbation by \eqref{eq:defect-pert-field}
is the $(1,2)$-defect, and so this is finally the situation which we
will study: 
\be
  \boxed{
  \parbox{0.85 \hsize}{ 
  \[ D_{12}(\laml\phi + \lamr\bar\phi)\]
    \text{The~(1,2)-defect~perturbed~by
  $\laml \phi + \lamr \bar\phi$
  ~with~$\phi = \phi_{h_{(1,3)},0}$~and
  $\bar\phi = \phi_{0,h_{(1,3)}}$.}}
  }
\labl{eq:MM-12-pert}
There are two useful results which we can derive for this
perturbed defect which we give now.

\subsubsection{Translation invariance}
\label{sss:trans}

As mentioned above, it is believed that a topological
defect perturbed by a chiral field still commutes with
$L_0+\bar L_0$ \cite{Konik:1997}. 
This is shown  in \cite{Runkel:2007wd} for chiral fields with $h<1/2$
where there are no UV divergences in the expansion of the perturbed
defect; for fields with $h\geq 1/2$ this property depends on the
existence of a suitable regulator which preserves the
commutators. Assuming for the moment that such a regulator can be found,
in the present case this would imply 
\be
   [L_0+\bar L_0, D_{(1,2)}(\lambda \phi)] = 0
   \quad \text{and} \quad
   [L_0+\bar L_0, D_{(1,2)}(\lambda \bar\phi)] = 0 ~.
\labl{eq:trans}
This guarantees that the action of $D_{(1,2)}(\lambda \phi)$ and
$D_{(1,2)}(\lambda \bar\phi)$ on a boundary state is non-singular. 

One particular consequence of the property \eqref{eq:trans} is that
the spectrum of states of the model on a strip containing a defect
parallel to the edges of the strip is independent of the position of
the defect.  This is easy to see in the case of the TCSA regulator, as
we prove in appendix \ref{app:TCSA-posn-indep}, so that it is very
reasonable to apply this assumption to our TCSA results and we will
use this in section \ref{ss:chiral} to deduce the end-points of the
chiral perturbations.

\subsubsection{Commutation with a subset of topological defects}
\label{sss:comm}

Let $D(\sum_\alpha \lamb{\alpha} \phi_\alpha)$ be a topological  
defect perturbed by defect fields $\phi_\alpha$ 
where $\phi_\alpha$ is a field (not necessarily primary) in the sector
$R_{k_\alpha}\otimes \bar R_{l_\alpha}$.
Suppose that either all of the $k_\alpha$ are of the form 
$(1,s_\alpha)$ with $s_\alpha$ odd, or all of the $l_\alpha$
are of the form $(1,s_\alpha)$ with $s_\alpha $ odd. Then
\be
   \big[\, D_{(r,1)} \,,\,D({\textstyle \sum_\alpha \lamb{\alpha}
\phi_\alpha})\,\big] = 0
   \quad \text{for}~1 \le r \le p{-}1~.
\labl{eq:comm}
We will demonstrate this in the case that the $k_\alpha$
are of the form $(1,s_\alpha)$ with $s_\alpha$ odd. The second case
can be seen analogously. 

Let $x \in R_i \otimes \bar R_i$ and $y \in R_j \otimes \bar R_j$ for
$i = (a,b)$ and $j = (c,d)$ elements of $\Ic_p$. From the explicit  
form of $D_{(r,1)}$ in \eqref{eq:bndstate-defectop} and abbreviating
$D \equiv D({\textstyle \sum_\alpha \lamb{\alpha} \phi_  \alpha})$ we
find 
\be
   \cev{x} D_{(r,1)} \, D \vec{y}
   = \frac{S_{(r,1)(a,b)}}{S_{(1,1)(a,b)}}
     \cev{x} D \vec{y}
   ~~,~~~
   \cev{x} D \, D_{(r,1)} \vec{y}
   = \frac{S_{(r,1)(c,d)}}{S_{(1,1)(c,d)}}
     \cev{x} D \vec{y} ~.
\labl{eq:comm-aux1}
Now expand out the exponential in $\cev{x} D \vec{y}$. Since each term  
in the integrand is 
in the sector $R_{(1,s_\alpha)}\otimes \bar R_{l_\alpha}$,
the expression $\cev{x} D \vec{y}$ can be non-zero   
only if $i$ and $j$ are such that $a{=}c$ and $b{-}d$ is
even. Substituting   the expression \eqref{eq:min-mod-S} for $S_{ij}$ gives
\be
   \frac{S_{(r,1)(a,b)}}{S_{(1,1)(a,b)}}
   = (-1)^{(r{+}1)b} \, 
  \frac{\sin( \pi (p{+}1) r a / p ) }
       {\sin( \pi (p{+}1) a   / p ) )} ~.
\ee
Since $a{=}c$ and $b{-}d$ is even, 
the two expressions in  
\eqref{eq:comm-aux1} are therefore equal for all states $x,y$, proving  
\eqref{eq:comm}.

\medskip

For us the commutator \eqref{eq:comm} is interesting because it implies
$[D_{(r,1)},D_{(1,2)}(\laml \phi + \lamr \bar\phi)] = 0$ with $\phi$  
and $\bar\phi$
as in \eqref{eq:MM-12-pert}. However,
let us note in passing that \eqref{eq:comm} can also be applied to  
bulk perturbations.
In particular, we can set $D = \exp(-\ell H)$ with $H$
the Hamiltonian
of the minimal model perturbed by the bulk field $\phi_{(1,3),(1,3)}$
since
\be
H  
  = \int \left(\;
\tfrac{1}{2\pi}({T {+} \bar T}) 
  + \mu\phi_{(1,3),(1,3)} \;\right)\;\mathrm{d}x
\labl{eq:Ham-bulk}
and $T$ and $\bar T$ are descendent fields in the $(1,1)$ representation.
The topological defects $D_{(r,1)}$ continue to commute with $D$ for  
all values of the coupling $\mu$,
and thus they are conserved charges also off criticality. This  
provides an independent argument
for the observation made in \cite{Fredenhagen:2009tn} that under  
the bulk flow between
two minimal models generated by $\phi_{(1,3),(1,3)}$,
the topological defect $D_{(r,1)}$ should again flow to a topological  
defect.

\sect{Exact results}\label{sec:exact}

The properties 
\eqref{eq:trans} and
\eqref{eq:comm}
of $D_{(1,2)}(\laml \phi + \lamr \bar\phi)$ enable us to find two
exact results for the perturbed defect, which we describe in sections
\ref{ss:chiral} and \ref{ss:fcir}. We also summarise in section
\ref{ss:ising} the known
results for the Ising model (the case $p=3$) which is distinct from
the other models.

\subsection{The chiral perturbations}
\label{ss:chiral}

The translation invariance of the purely chiral and purely anti-chiral
perturbations in \erf{eq:trans} allows us to determine the nature of
these flows exactly.

Consider the situation described in \eqref{eq:MM-12-pert}, 
i.e.\ $D_{(1,2)}(\laml \phi + \lamr \bar\phi)$ is
the operator obtained by placing the $(1,2)$-defect on the unit circle
and perturbing by $\laml \phi + \lamr \bar\phi$ with $\laml$ and
$\lamr \in \mathbb{R}$. Let $\psi$ be the
primary boundary field on the $(1,2)$-boundary condition in the
$R_{(1,3)}$ representation and denote by $\vvecc{(1,2)+\lambda \psi}$
the boundary state obtained by perturbing the $(1,2)$-boundary by
$\lambda \psi$. 
One
can normalise the fields $\phi$, $\bar\phi$ and $\psi$ such that the
identity\footnote{\label{fn:1}
  This identity is evident in the case $\lambda=0$. Since the space of
  $(1,3)$-primary fields on the $(1,2)$-boundary is one-dimensional
  and spanned by $\psi$, the perturbing fields $\phi$ and $\bar\phi$
  have to be proportional to $\psi$ once the $(1,2)$-defect is pushed
  on top of the $(1,1)$-boundary. Alternatively, the identity can be
  proved with the methods in \cite{Frohlich:2006ch}.} 
\be
  D_{(1,2)}(\lambda \phi) \vvecc{1,1}
  = \vvecc{(1,2)+\lambda \psi}
  = D_{(1,2)}(\lambda \bar\phi) \vvecc{1,1} 
\labl{eq:12-def-12-bnd}
holds. The boundary flows in the middle of this series of equalities are already known \cite{Lesage:1998qf,Recknagel:2000ri,Graham:2000si},
\be
  \vvecc{1,1} 
  ~\xleftarrow{~~ -\infty \leftarrow \lambda ~~}~ 
  \vvecc{(1,2)+\lambda \psi} 
  ~\xrightarrow{~~\lambda \rightarrow +\infty~~}~  
  \vvecc{2,1} ~.
\labl{eq:12-bnd-flow}
        As mentioned in section \ref{sec:chirdefpert},
a chirally perturbed topological
defect necessarily has again a topological defect as an IR fixed
point. Acting with the IR topological defect on $\vvecc{1,1}$ gives a
conformal boundary state that has to agree with
\eqref{eq:12-bnd-flow}. The explicit expressions in
\eqref{eq:bndstate-defectop} show that a topological defect is
uniquely determined by its action on $\vvecc{1,1}$, so that the above
reasoning fixes the horizontal and vertical flows in figure
\ref{fig:flow-p>3} to be 
\be
\begin{array}{rcccl}
  I \equiv D_{(1,1)}
&  ~\xleftarrow{~~ -\infty \leftarrow \laml ~~}~ 
&  D_{(1,2)}(\laml\phi)
&  ~\xrightarrow{~~\laml \rightarrow +\infty~~}~  
&  D' \equiv D_{(2,1)}
\;,
\\
  I \equiv D_{(1,1)}
&  ~\xleftarrow{~~ -\infty \leftarrow \lamr ~~}~ 
&  D_{(1,2)}(\lamr\bar\phi)
&  ~\xrightarrow{~~\lamr \rightarrow +\infty~~}~  
&  D' \equiv D_{(2,1)}
\;.
\end{array}
\labl{eq:12-def-flow}

\subsection{The factorising component of any IR fixed point}
\label{ss:fcir}

The fact that our perturbed defects commute with the set of
topological defects $\{D_{(t,1)}\}$ shown in \erf{eq:comm} allows us
to determine the building blocks of any factorising component of any  
IR fixed
point.

Consider the superposition of
elementary factorising defects given by
\be
   F^{s|s'} = \sum_{t=1}^{p-1} D_{(t,1)} \vvecc{1,s}\ccevv{1,s'} D_{(t, 
1)}
   \cdot \begin{cases}
   \tfrac12 &: p \text{ odd and } s=s'=\tfrac{p+1}2 \\
   1 &: \text{otherwise}
   \end{cases} \quad .
\labl{eq:Fss-def}
The factor of $\tfrac12$ has to be included because for $s=s'=\tfrac{p 
+1}2$ we have
$\vvecc{t,\tfrac{p{+}1}2}\ccevv{t,\tfrac{p{+}1}2} = \vvecc{p-t, 
\tfrac{p{+}1}2}\ccevv{p-t,\tfrac{p{+}1}2}$,
so that in this case each factorising defect appears twice in the sum.  
For example, in the Ising model ($p=3$)
we have $F^{2|2} = \vvecc{1,2}\ccevv{1,2}$.

It is shown in appendix \ref{app:Dr1-orbit} that $[D_{(r,1)},F^{s| 
s'}]=0$ for $r=1,\dots,p{-}1$, and that the following statement is true:
If a conformal defect $C$ obeys $[D_{(2,1)},C]=0$ (which implies that $ 
[D_{(r,1)},C]=0$ for all $r$) then we can write
\be
   C = R + F
\labl{eq:C-fact-rest}
where $R$ is a conformal defect that does not contain factorising
defects as summands, i.e.\ $(R,\vvecc{a}\ccevv{b})=0$ for all $a,b \in
\Ic_p$, and where $F$ is a combination of factorising defects taking
the form
\be
   F = \sum_{s,s'=1}^p \sum_{a=1}^{p-1} m_{a}^{s|s'}
   D_{(a,1)} F^{s|s'}
\labl{eq:C-fact-rest2}
for suitable constants $m_{a}^{s|s'} \in \Zb_{\ge 0}$.
If $C$ is an IR fixed point of $D_{(1,2)}(\laml \phi +
\lamr \bar\phi)$ in a particular direction in the
$(\laml,\lamr)$-plane, it will commute with the $D_{(r,1)}$ and hence
can be written in the form \eqref{eq:C-fact-rest}.
This will be exploited in the $g$-function analysis in section
\ref{sec:pert} and in identifying the spectrum on a strip with a
perturbed defect via TCSA in section \ref{sec:TCSA-results}.

\subsection{The Ising case}
\label{ss:ising}

The Ising model is unique amongst unitary minimal models in that its
conformal defects can be completely classified. 
The reason is that its square Ising$\otimes$Ising is a $c=1$ CFT which
can be described in terms of an orbifolded free 
boson. Using this relation, Oshikawa and Affleck 
give a complete classification in \cite{Oshikawa:1996dj} of the
conformal defects of the Ising model in terms of the conformal
boundary conditions of the orbifolded boson as follows.

There is a continuous family of Dirichlet boundary conditions
$D(\fii_0)$ 
subject to 
$D(\fii_0)=D(-\fii_0)=D(\fii_0 + 2\pi)$ for which a fundamental domain
is $\fii_0\in[0,\pi]$;
and a
continuous family of Neumann boundary conditions $N(\tilde\fii_0)$ 
with
$N(\tilde\fii_0)=N(-\tilde\fii_0)=N(\tilde\fii_0 + \pi)$ for which a
fundamental domain is $\tilde\fii\in[0,\frac\pi2]$.
The boundary conditions in the two continuous families are all
elementary, except for the boundary conditions at the endpoints of the
fundamental domains, which each split into two elementary 
boundary conditions so giving 
a discrete set of eight
boundary conditions. Altogether we have
\begin{align*}
\vec{D(\fii_0)}\quad&\text{with }\fii_0\in(0,\pi)\quad\text{and
}\vec{D(0)}_\pm\,,\;\vec{D(\pi)}_\pm\,,\\ 
\vec{N(\tilde\fii_0)}\quad&\text{with }\tilde\fii_0\in(0,\pi/2)\quad\text{and
}\vec{N(0)}_\pm\,,\;\vec{N(\pi/2)}_\pm\,.
\end{align*}
These have $g$ values $\sqrt 2$ for the continuous Neumann series,
$1$ for the continuous Dirichlet series, $1/\sqrt 2$ for the discrete
Neumann defects and $1/2$ for the discrete Dirichlet defects.

\begin{figure}

\setlength{\unitlength}{4400sp}%
\begingroup\makeatletter\ifx\SetFigFont\undefined%
\gdef\SetFigFont#1#2#3#4#5{%
  \reset@font\fontsize{#1}{#2pt}%
  \fontfamily{#3}\fontseries{#4}\fontshape{#5}%
  \selectfont}%
\fi\endgroup%
\begin{picture}(5200,1728)(2766,-4630)
\put(4988,-4411){\circle*{76}}
\put(4988,-4411){\makebox(0,0)[rc]{$(f-)=N(0)_-$~~}}
\put(4988,-4111){\circle*{76}}
\put(4988,-4111){\makebox(0,0)[rc]{$(f+)=N(0)_+$~~}}
\put(7013,-4111){\circle*{76}}
\put(7013,-4411){\circle*{76}}
\put(7013,-4111){\makebox(0,0)[lc]{~~$(+f)=N(\frac\pi 2)_+$}}
\put(7013,-4411){\makebox(0,0)[lc]{~~$(-f)=N(\frac\pi 2)_-$}}
\put(6001,-4186){\line( 0,-1){150}}
\put(6001,-4156){\makebox(0,0)[cb]{$\sigma=N(\frac\pi 4)$}}
{\thicklines
\put(5101,-4261){\line( 1, 0){1800}}
}
\put(4088,-3511){\circle*{76}}
\put(4088,-3511){\makebox(0,0)[rc]{$(--)=D(0)_-$~~}}
\put(4088,-3211){\circle*{76}}
\put(4088,-3211){\makebox(0,0)[rc]{$(++) = D(0)_+$~~}}
\put(7913,-3211){\circle*{76}}
\put(7913,-3511){\circle*{76}}
\put(7913,-3211){\makebox(0,0)[lc]{~~$(-+)=D(\pi)_+$}}
\put(7913,-3511){\makebox(0,0)[lc]{~~$(+-)=D(\pi)_-$}}
{\thicklines
\put(4201,-3361){\line( 1, 0){3600}}
}
\thinlines
\put(6001,-3286){\line( 0,-1){150}}
\put(6001,-3506){\makebox(0,0)[ct]{$(ff)=D(\frac\pi 2)$}}
\put(5101,-3286){\line( 0,-1){150}}
\put(5101,-3256){\makebox(0,0)[cb]{$1=D(\frac\pi 4)$}}
\put(6901,-3286){\line( 0,-1){150}}
\put(6901,-3256){\makebox(0,0)[cb]{$\epsilon=D(\frac{3\pi}4)$}}
\end{picture}%

\caption{The Dirichlet (top) and Neumann (bottom) boundary conditions on the
  orbifolded free boson and the corresponding topological and
  factorising defects in the folded Ising model  } 

\label{fig:ising-bcs}
\end{figure}
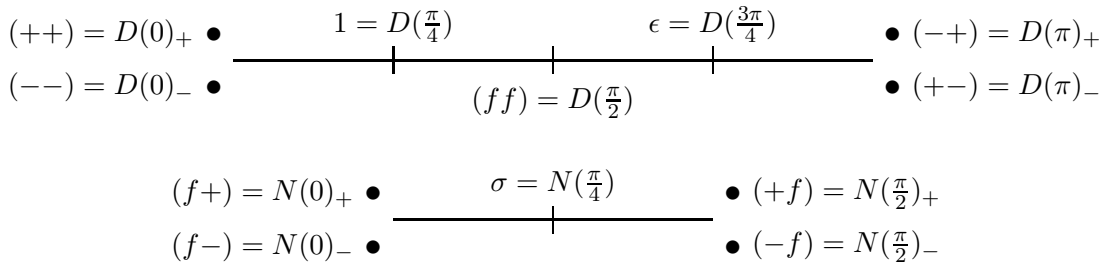

There are three topological defects in the Ising model given by Kac
labels $(1,1)=\id$, $(1,2)=\sigma$ and $(2,1)=\eps$ and three Cardy
boundary conditions ``$+$''$=(1,1)$, ``$f$''$=(1,2)$ and
``$-$''$=(2,1)$. The topological and factorising defects are
identified with the orbifolded boson boundary conditions 
\cite{Oshikawa:1996dj,Quella:2006de} as shown in figure
\ref{fig:ising-bcs}.

To identify the fixed points of the numerical RG flows in section
\ref{sec:Ising} we need the partition function encoding the spectrum
of states on a strip with the $(1,1)=$``$+$'' boundary condition on
the two edges and a defect in the middle. When this strip is folded,
this is equal to the partition function in the orbifolded boson model
with the $(++)=D(0)_+$ boundary condition at one side of the folded
strip and some boundary condition corresponding to the Ising defect at
the other boundary. In the UV 
this is the $(1,2)=\sigma=N(\pi/4)$ boundary condition, but by the
$g$-theorem, the IR point fixed point can be any Dirichlet boundary
condition (together with any of the four discrete Neumann conditions,
but these do not arise in the flows we consider); the partition
function of such a system is given in 
\cite{Oshikawa:1996dj}
\be
Z_{D(0)_+,D(\fii_0)}(q;\frac12)
=\frac{q^{\frac12(\frac{\fii_0}\pi)^2}}{\eta(q)}
\sum_{n=-\infty}^\infty q^{2n^2+2n\frac{\fii_0}\pi}\,,
\labl{eq:Z.OA}
where $q=e^{-2\pi L/R}$. This expression is also valid at the
endpoints $\fii_0=0,\pi$ if one takes $D(0)=D(0)_++D(0)_-$ and
$D(\pi)=D(\pi)_++D(\pi)_-$. 

As has been discussed in section \ref{ss:chiral}, the IR
fixed points of the purely chiral flows can be easily found: they are
the $(2,1)=\eps$ and $(1,1)=\id$ topological defects in the positive
and negative directions, respectively.
The remaining fixed points have been recently found (amongst other
results) in \cite{Fendley:2009gm} and the space of flows is shown in figure
\ref{fig:finalflows}(a).  
In that paper, the perturbed defects are parametrised by an angle
$\theta$ with the general result
\be
N(\tilde\fii_0)\;\longrightarrow\; D(\tilde\fii_0 + \theta)
\;.
\labl{eq:FFNresult1}
We define the angle $\alpha$ by $\tan(\alpha)=\lamr/\laml$,
with 
$\alpha \in (-\tfrac\pi2,\tfrac\pi2)$ for $\laml>0$ and 
$\alpha \in (\tfrac\pi2,\tfrac{3\pi}2)$ for $\laml<0$.
One can check that $\alpha$ is related to $\theta$ by $\alpha=\pi-\theta$
so that the 
prediction of \cite{Fendley:2009gm} for the RG flows of the defects 
\eqref{eq:MM-12-pert}
is
\be
D_\sigma = N(\pi/4)\;\longrightarrow\; D(5\pi/4 - \alpha)
\;,
\labl{eq:FFNresult2}
which is exactly what we shall find in section \ref{sec:Ising}.

\sect{Perturbative analysis}\label{sec:pert}

The renormalisation group equations for perturbations of conformal
boundary conditions (and hence also of defects) have been known for a
long time, studied first in \cite{Affleck:1991tk} and used
in \cite{Recknagel:2000ri,Graham:2001pp} to study the flows
in unitary minimal models.

The necessary ingredients are a conformal boundary condition, a 
set of relevant boundary fields $S=\{\phi_i\}$ which is closed in the sense
that all the relevant fields occurring in the fusion of any two fields
in $S$ is also in $S$, the conformal dimensions $h_i$ of these fields,
and the structure constants $c_{ijk}$ between these relevant fields. 
Furthermore, since the perturbative integrals are divergent for
$h_i\geq 1/2$, it is necessary to regularise them to which end we
introduce a UV cut-off $a$ as in \cite{Affleck:1991tk} and define
dimensionless couplings $\mu_i = \lamb{i} a^{y_i}$ where $y_i = 1 -
h_i$. 

Using the regularisation in \cite{Affleck:1991tk}, the RG equations
for the couplings $\mu_i$  are
\be
  \dot\mu_i = y_i\mu_i - \sum_{j,k} c_{ijk}\mu_j\mu_k + O(\mu^3)
\;.
\ee
The change in the $g$-value of the boundary condition/defect is also
known in perturbation 
theory \cite{Affleck:1991tk,Recknagel:2000ri} and is (to third order)
\be
  \log( g(\mu_i) )
- \log(g_0)
= -\pi^2 y_i \mu_i^2 + \frac{2\pi^2}{3} \sum_{j,k}
  c_{ijk}\mu_i\mu_j\mu_k 
\;.
\ee
The scale dimensions of the perturbing fields at a fixed point can
be read off by linearising the RG equations about the new fixed
point, the eigenvalues being $1{-}h'$ where $h'$ is the scale
dimension of the field \cite{Graham:2001pp}.

We now restrict attention to the perturbed defects
\erf{eq:MM-12-pert}. In this case the space of relevant defect fields
generated by $\phi$ and $\bar\phi$ closes on those fields alone, that
is they generate no new relevant defect fields, so we do not need to
introduce couplings to any other fields into the RG equations.

Secondly, since $\phi$ and $\bar\phi$ are chiral and anti-chiral respectively,
the only structure constants appearing in the RG equations which can
be non-zero\footnote{%
It is important to note that $\scc\neq 0$
for $p>3$ but when $p=3$ the constant $\scc$ vanishes, so that this case
is not directly amenable to the analysis here.}
are $c_{\phi\phi\phi}$ and 
$c_{\bar\phi\bar\phi\bar\phi}$, and these are in fact equal and the
same as the boundary structure constant $\scc$ for the
field $\psi\equiv\psi_{(1,3)}$ on the $(1,2)$ conformal boundary condition
(see footnote \ref{fn:1}
and \cite[sect.\,2]{Runkel:2007wd}). 
This means
that the defect RG equations are two decoupled copies of the boundary RG
equations of \cite{Recknagel:2000ri}, that is
\bea
 \dot\mul = y\mul - \scc \mul^2 \;, \\
 \dot\mur = y\mur - \scc \mur^2
\;,
\eear\labl{eq:pflows}
where $y = 1 - h_{1,3} = \frac{2}{p+1}$ and $\scc=\sqrt{8/3} + O(y)$.
The $g$ value of the perturbed defect is given to third order by
\be
 \log(g(\mul,\mur)) - \log(g(0,0))
= - \pi^2 y (\mul^2 + \mur^2)
  + \frac{2 \pi^2}{3} \scc (\mul^3 + \mur^3)
\;.
\ee
Each RG equation has two fixed points, $\mu=0$ and
$\mu=\mu^*$ (where $\mu^*=y/\scc$) so that the combined RG
equations have four fixed points. 

It is important to note that the perturbation expansion is for small
$y$, that is for large $p$ where the central charge is close to 1. 
While the perturbative results for boundary flows in \cite{Recknagel:2000ri}
still hold true all
the way down to $p=3$, this will not be the case here, as the spectrum
of the IR fixed points is quite different at $p=3$.

\begin{figure}[tb]

$$ 
  \begin{picture}(160,180)
  \put(0,0){\scalebox{0.60}{\includegraphics{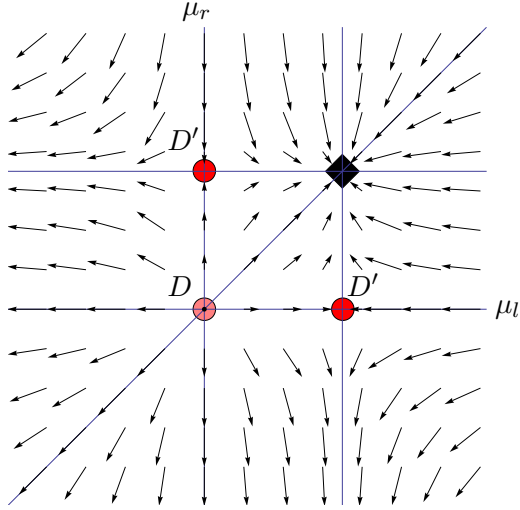}}}
  \put(0,0){
     \setlength{\unitlength}{.80pt}\put(-90,-587){
     \put(320,678)   {$\mul$}
     \put(172,819)   {$\mur$}
     \put(165,755)   {$D'$}
     \put(165,686)   {$D$}
     \put(250,686)   {$D'$}
     }\setlength{\unitlength}{1pt}}
  \end{picture}
$$
\caption{The perturbative flows for the system \eqref{eq:pflows}.
The perturbative fixed points are as in figure \ref{fig:flow-p>3},
the black diamond being the conformal defect $C$.}
\label{fig:pflow}
\end{figure}

The perturbative RG flows are shown in figure
\ref{fig:pflow}.

The values of 
the couplings at the fixed points, the corresponding value of $g$
and the scale dimensions of the perturbing fields at the fixed point
(to first order in $y$)
are given below
\be
 \begin{array}{c|c@{~~}cc}
(\mul,\mur) & \log(g/g_0) & h_\phi & h_{\bar\phi} \\
\hline
(0,0)     &   0               & 1-y & 1-y\\
({\mu^*},0)   & - \frac{\pi^2 y^3}8   & 1+y & 1-y\\
(0,{\mu^*})   & - \frac{\pi^2 y^3}8  & 1-y & 1+y\\
({\mu^*},{\mu^*}) & - \frac{\pi^2 y^3}4 & 1+y & 1+y\\
\end{array}
\labl{eq:fps}
The topological defects of lowest $g$ value and their
expansions in $y$ to third order are
\be
 \begin{array}{c|cl}
\hbox{Defect} & g & \log(g) \\
\hline
D_{(1,1)},D_{(p-1,1)}  & 1 & 0 \\
D_{(2,1)},D_{(p-2,1)}  & 2 \cos(\frac{\pi}{p})   & \log(2) - \frac{\pi^2
  y^2}8 - \frac{\pi^2 y^3}8 \\
D_{(1,2)},D_{(1,p-1)}  & 2 \cos(\frac{\pi}{p+1}) & \log(2) - \frac{\pi^2
  y^2}8  \\
D_{(3,1)},D_{(p-3,1)}  & 1 + 2 \cos(\frac{2\pi}{p})   & \log(3) - \frac{\pi^2
  y^2}3 - \frac{\pi^2 y^3}3 \\
D_{(1,3)},D_{(1,p-2)}  & 1 + 2 \cos(\frac{2\pi}{p+1}) & \log(3) - \frac{\pi^2
  y^2}3  \\
\end{array}
\ee
Comparing $g$-values and noting that the end point of an RG flow
generated by a chiral field must be topological, we see that the
chiral and anti-chiral perturbations of ${D_{(1,2)}}$ can a priori
have either the $D_{(2,1)}$ or $D_{(p-2,1)}$ defect as their
perturbative fixed point. This can be decided by considering two point
functions of bulk fields in the presence of the defect, which can be
calculated using \eqref{eq:bndstate-defectop}:
\be\renewcommand{\arraystretch}{1.2}
\begin{array}{l|ccc}
 & D_{(1,2)} & D_{(2,1)} & D_{(p-2,1)} \\
\hline
\langle \varphi_{(12,12)} | D | \varphi_{(12,12)} \rangle
& 
- 2 \cos\frac{2\pi}{p+1}
& 
- 2 \cos\frac{\pi}{p}
&
- 2 (-1)^p \cos\frac{\pi}{p}
\\
\langle \varphi_{(21,21)} | D | \varphi_{(21,21)} \rangle
& 
- 2 \cos\frac{\pi}{p+1}
& 
- 2 \cos\frac{2\pi}{p}
&
- 2 (-1)^{p+1} \cos\frac{2\pi}{p}
\\
\end{array}
\labl{eq:two-pts}
We see that the two-point functions in the
presence of $D_{(2,1)}$ are perturbatively close to those in the
presence of $D_{(1,2)}$,
whereas in the presence of $D_{(p-2,1)}$ the two point functions of 
$\varphi_{(12,12)}$ and 
$\varphi_{(21,21)}$ are not perturbatively close for $p$ odd and $p$
even respectively.
Thus we deduce that the perturbative fixed point for the chiral
perturbations of $D_{(1,2)}$ are $D_{(2,1)}$ and not $D_{(p-2,1)}$.

Equally, we see that the $C$ defect which is the fourth fixed point in
the `++' direction is not a topological defect.
The question remains whether $C$ can be found as a superposition of
factorising defects or the Identity defect superposed with a set of
factorising defects. 

This is easy to address given the general form of a fixed point in
\eqref{eq:C-fact-rest}.
    We first compute the $g$-value of each summand in \eqref{eq:C-fact-rest2}
\begin{eqnarray}
g(  D_{(a,1)} F^{s|s'} )
&=&
 \left(\frac{\sin(\frac{a\pi}{p})}{\sin(\frac{\pi}{p})}\right)
 \left(\frac{\sin(\frac{s\pi}{p+1})}{\sin(\frac{\pi}{p+1})}\right)
 \left(\frac{\sin(\frac{s'\pi}{p+1})}{\sin(\frac{\pi}{p+1})}\right)
 \cdot
\left(
\sqrt{\frac{2p}{p+1}}\, \frac{\sin(\frac{\pi}{p+1})}{\sin(\frac{\pi}{p})}
\right) \cdot \delta
\nonumber\\
&=&
a s s' \left(\sqrt 2 - \frac{3}{2\sqrt 2}y\right) + O(y^2)
\;.
\end{eqnarray}
  Here $\delta$ is either $\tfrac12$ or $1$ as in \eqref{eq:Fss-def}; it does not appear 
  in the second line as this is evaluated for fixed $s,s'$ and large $p$.

Since each of the combinations $D_{(a,1)}F^{s|s'}$ has a $g$-value
greater than or equal to one as does each of the topological defects,
and since the $g$-value of $D_{(1,2)}$ is less than two, it is clear
that any IR fixed point of a flow starting from $D_{(1,2)}$ can
include at most one topological defect or one factorising combination
but not both. 
Furthermore, for large $p$, the only possible
factorising combinations with small enough $g$ value are $F^{1|1}$ and
$D_{(p-1,1)}F^{1|1}$ but these do not agree with the perturbative
calculation of the $g$-value of $C$.

(We note here that there is an additional candidate in the case $p=4$:
the factorising defect $D_{(2,1)}F^{1|1}$ also has a $g$-value lower
than that of $D_{(1,2)}$).

Summarising, from an analysis of $g$-values, $C$ is not a linear
superposition of topological defects and factorising defects and so
contains a new conformal defect; the most likely result is that $C$ is
a new elementary conformal defect, but that cannot be determined from
the calculations of the $g$ value alone.

There remains the question of identifying the nature of the fields
$\phi$ and $\bar\phi$ at the fixed points. 
We can do this by checking which fields on $D_{(2,1)}$ have weights as
given in table \eqref{eq:fps}. This is straightforward and  the only
candidates are as given below 
\be
 \begin{array}{c|c@{~~}cc}
(\mul,\mur) & D & \phi & {\bar\phi} \\
\hline
(0,0)     &   {D_{(1,2)}} & \phi_{(13)(11)} & \phi_{(11)(13)}\\
({\mu^*},0)   &   {D_{(2,1)}} & \phi_{(31)(11)} & \phi_{(33)(13)}\\
(0,{\mu^*})   &   {D_{(2,1)}} & \phi_{(13)(33)} & \phi_{(11)(31)}\\
({\mu^*},{\mu^*}) &   C    &        ?        &      ?         \\
\end{array}
\labl{tab:weights}
This means that we should be able to obtain the `C' defect by TCSA 
calculations starting both from $D_{(1,2)}$ and perturbing by
$\phi+\bar\phi$ and also starting from $D_{(2,1)}$ and perturbing by
$\phi_{(13,33)}$ (or equivalently by $\phi_{(33,13)}$). This is
one of the subjects of section \ref{sec:TCSA-results}.

\sect{Truncated conformal space approach for defects}
\label{sec:TCSA}

\subsection{The TCSA Hamiltonian for defects}
\label{ssec:TCSA-Ham}

The so-called truncated conformal space approach or TCSA is a
numerical method for calculating the spectrum of a perturbed conformal
field theory. One chooses a system which admits a Hamiltonian
description and then the Hamiltonian is restricted to a finite dimensional
subspace of the infinite dimensional Hilbert space spanned by energy
eigenstates whose energy eigenvalues (or conformal weights) are not
greater than a threshold value. The original idea was
proposed in \cite{Yurov:1989yu} and it was applied for the first time for
boundary problems in \cite{Dorey:1997yg}.

One can envisage several possible systems involving defects which allow such a
Hamiltonian description. We choose to apply a generalisation of the
boundary TCSA of \cite{Dorey:1997yg}, 
because the space of states is much smaller than in the other
possibilities (they form irreducible representations of a single copy of the
Virasoro algebra -- see \erf{eq:strip-states}). This makes the numerical
calculations easier and we can also make use of the known properties of the
interactions between defects and boundaries. The simplest situation is given
by a strip of width $R$ with conformal boundary condition $(1,1)$ on both
sides and a defect labelled by $k \in \Ic_p$ running parallel to the
boundaries at a distance $a$ from one boundary. According to
\erf{eq:strip-states} the Hilbert space consists of a sole representation
(independently of $a$),
\be 
 \Hc^{(1,1),(1,1)}_k = R_k ~.  
\ee

\begin{figure}[t].

$$  
  \begin{picture}(320,80)
  \put(0,0){\scalebox{1.20}{\includegraphics{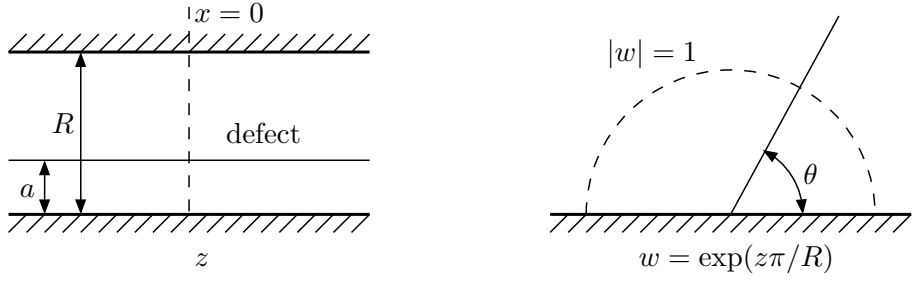}}}
  \put(0,0){
     \setlength{\unitlength}{1.20pt}\put(0,0){
     \put(5,12)   {$a$}
     \put(15,32)  {$R$}
     \put(60,67)   {$x=0$}
     \put(70,29)   {defect}
     \put(60,-10)   {$z$}
     \put(200,-10)   {$w = \exp(z \pi/R)$}
     \put(252,16) {$\theta$}
     \put(190,55) {$|w|=1$}
     }\setlength{\unitlength}{1pt}}
  \end{picture}
$$

\caption{\small The strip is mapped to the upper half-plane and a
  defect running parallel to the edges of the strip is mapped into a
  defect running in a radial straight line. The dotted line shows an
  equal time slice for the Hamiltonian description.} 
\label{fig:tcsa-setup}
\end{figure}

By the exponential map the strip unfolds onto the upper half plane
with the boundaries lying along the negative and the positive real
axis and the defect line running from the origin to infinity at an
angle $\theta=a \pi/R$, as shown in figure \ref{fig:tcsa-setup}.

The Hamiltonian of this conformal field theory perturbed by $N$ defect
fields is 
\be
H=\frac\pi{R}\left[L_0-\frac{c}{24}+
  \sum_{i=1}^N\lamb{i}\,
  \left(\frac{R}\pi\right)^{1-(h^i_{l}+h^i_{r})}\,
  e^{i(h^i_{l}-h^i_{r})\theta}
  \phi_{h^i_{l},h^i_{r}}(e^{i\theta},e^{-i\theta})\right]\,.
\labl{eq:pert-Ham}

We use a non-orthonormal basis consisting of vectors of the form
\be
L_{-n_1}\dots L_{-n_m}\vec{h_k}\,,\qquad
n_1\ge\dots\ge n_m>0\;.
\labl{eq:basis}
We label these basis vectors $\vec{v_i}$, their $L_0$ eigenvalues by
$\Delta_i$ and their inner product matrix $G_{ij}=\langle v_i
\vec{v_j}$.
The inner product matrix appears explicitly in the matrix to
be diagonalised: 
\be
h_{jk}=\frac\pi{r}\left[(\Delta_j-\frac{c}{24})\delta_{jk}+
 \sum_{i=1}^N \kapp{i}\,
  {e^{i(h^i_{l}-h^i_{r})\theta}}
  \,
  (G^{-1}B(\theta)^i)_{jk}\right]\,,
\ee
where we have introduced an energy scale $\Lambda$ so that the
operator $h = H/\Lambda$ is dimensionless and its matrix elements are
$h_{ij}$ where $h \vec{v_j} = h_{ij} \vec{v_i}$; furthermore,
$B(\theta)^i_{jk}$ are the matrix elements  
$\langle{v_j} | \phi_{h^i_{l},h^i_{r}}(e^{i\theta},e^{-i\theta}) 
 \vec{v_k}$ in the basis \erf{eq:basis}, 
$r = \Lambda R$ is a dimensionless parameter and 
$\kapp{i}=\lamb{i} (R/\pi)^{1-h^i_l - h^i_r}$
are dimensionless  coupling constants.

In the general case it is necessary to determine the structure
constants to normalise the matrix elements $B(\theta)^i$ correctly,
but in the cases we consider the Hilbert space consists of a single
representation so that there is a single highest weight $|h_k\rangle$
and so we normalise the fields so that 
$\langle h_k | \phi_{h_i,\bar h_j}(1) | h_k \rangle = 1$.

The two special cases we will consider are the case $N=1$ of a single 
non-chiral perturbation and the case $N=2$ where the perturbing fields
are two chiral perturbations 
\erf{eq:defect-pert-field} with $(h^1_l,h^1_r)=(h,0)$ and
$(h^2_l,h^2_r)=(0,h)$. 
In the first case, there is a single dimensionless
coupling constant $\kappa$ and the Hamiltonian becomes
\be
  \frac{r}{\pi}\,
  h_{jk}=(\Delta_j-\frac{c}{24})\delta_{jk}+
  \kappa\,
  e^{i(h_{l}-h_{r})\theta}\,
  (G^{-1}B(\theta))_{jk}
\,.
\ee
In the second case, the two coupling constants are $\kapl = \kapp{1} =
\kappa\cos(\alpha) $ and $ \kapr = \kapp{2} =\kappa\sin(\alpha)$ so that 
\be
  \frac{r}{\pi}
  h_{jk}
  =(\Delta_j-\frac{c}{24})\delta_{jk}+
  \kappa\,\left(
  \cos(\alpha)\,e^{ih\theta}\,
  (G^{-1}B(\theta))_{jk}
\;+\;
  \sin(\alpha)\,e^{-ih\theta}\,
  (G^{-1}B(\theta)^*)_{jk}
\;
 \right)
\,,
\labl{eq:two-k-ham}
where $B(\theta)^*$ is the complex conjugate of the matrix $B(\theta)$.

From the equality
\be
    (\Delta_i - \Delta_j)\3pt{v_i}{\phi_h(z)}{v_j}
=   \3pt{v_i}{[L_0,\phi_h(z)]}{v_j}
  = \left(h+z\frac\partial{\partial z}\right)
    \3pt{v_i}{\phi(z)}{v_j}
\;,
\ee
it follows that the matrix elements of a chiral field between $L_0$
eigenstates have coordinate dependence 
\be
\3pt{v_i}{\phi_h(z)}{v_j}=C\cdot z^{\Delta_i - h - \Delta_j}\,.
\labl{eq:3pt-coord-dep}
This implies that
\be
e^{ih\theta}B(\theta)_{kl}+e^{-ih\theta}B(\theta)^*_{kl} \propto
e^{i(\Delta_k-\Delta_l)\theta}+e^{i(\Delta_l-\Delta_k)\theta}\,.
\ee
When the defect is in the middle of the strip, $\theta=\pi/2$, and the
couplings to the left and right fields are the same
($\cos(\alpha)=\sin(\alpha)$) then the
$(k,l)$ matrix element of the perturbing Hamiltonian is proportional to
$i^{\Delta_k-\Delta_l}+(-i)^{\Delta_l-\Delta_k}$. If
$\Delta_l-\Delta_k$ is an odd integer, the two terms cancel each other
and the matrix element becomes exactly zero. 
In all our examples, the Hilbert space consists of a single representation, so
the weight differences are always integers and they are odd if one of
the vectors belongs to an even level while the other one belongs to an
odd level. This means that for $\theta=\pi/2$ and $\kapl=\kapr$, the
even and odd levels of the representation constitute two disconnected
sectors: the Hamiltonian has zero matrix elements between vectors
belonging to different sectors and thus it is block diagonal.

\subsection{Finite-Size scaling in TCSA}
\label{sec:fss}

There are two RG--type flows at play in the TCSA system. The first is the
physical flow in which we are interested, that is the flow induced by
scaling the size of the system. We parametrise this flow by a
parameter $t$. The second is that induced by changing
the cutoff $N$ limiting the size of the space of states. 
We are interested in the case of two relevant fields with a
Hamiltonian \eqref{eq:two-k-ham} in which case there are three couplings
to consider, $r(N,t)$ and $\kapp{i}(N,t)$.

Considering the case of  infinite cut-off first, we expect that 
the TCSA system approaches the system with canonical scaling, so that
at $N=\infty$
\be
\begin{cases} &\dot r = r \\ &\dot{\kapp{i}} = y \kapp{i}
\end{cases}
\;\;\hbox{ or }\;\;
\begin{cases} &r = r_0 e^t \\ &\kapp{i} = \kappp{i}{0}\, e^{yt}
\end{cases}
\ee
For finite $N$, these are both altered. 
This has two effects: firstly,
the eigenvalues of $h_{jk}$ are related to those of the renormalised
Hamiltonian up to an unknown factor, so that it is only ratios of
energy differences that can be calculated accurately; secondly,
the TCSA flow can approximate IR fixed points for finite values of the bare
coupling constants or equivalently a finite value of the bare volume
\cite{Feverati:2006ni}, so that we can find one, or indeed more, conformal
field points for increasing values of $\kappa$.

We can find the change in the coupling constants as $N$ changes,
which from \cite{Feverati:2006ni} is according to an RG-type equation
of the form
\be
 - N \frac{\partial \kapp{i}}{\partial N}(N,t)
 = \bt{i}(\kappa(N,t);N)
\;,
\ee
where
\be
 \bt{i}(\kappa;N)
= - c_{ijk}\kapp{j}\kapp{k} N^{-y} 
\;+\; o(\kappa^{2},N^{-y},y)
\;.
\ee
We can use this equation to find the finite-size scaling flow at fixed
cut-off under two assumptions.

The first assumption, already stated, is that in the $N\to\infty$
limit, the TCSA couplings become the ``linear beta function''
couplings, a fact assumed in much of the literature, so that the
finite-size flow at $N=\infty$ is 
\be
 \kapp{i}(\infty,t) = e^{yt}\;\kappp{i}{0}
\;.
\ee
The second assumption is that the leading terms in the beta-functions
take a simple scaling form,
that is
\be
  \bt{i}(\kappa;N) = N^y \gamm{i}(\kappa N^{-y}) + \ldots
\;.
\labl{eq:betat}
This is true for the quadratic terms and a reasonable assumption for
the leading behaviour of the higher order terms.

Given these two assumptions, we can then deduce the following three
results (see appendix \ref{app:TCSA})
\begin{eqnarray}
    \kapp{i}(N e^t,t) 
&=& e^{yt}\, \updown{f}{i}(\kappp{}{0})
\label{eq:fss1}
\\
    \kapp{i}(N , t) 
&=& e^{yt}\,\kapp{i}(N e^{-t},0)
\label{eq:fss2}
\\
    \frac{\partial}{\partial t}\kapp{i}(N,t) 
&=& \bb{i}( \kappa(N,t) ; N)
\label{eq:fss3}
\end{eqnarray}
The first result states that we can follow a finite-size scaling flow
by scaling the couplings by the same factor but also increasing the
cut-off. This is useful if we do not know the beta-functions, as is
the case here. 
The second result gives the scaling flow at fixed $N$ in terms of the
couplings at smaller $N$, and the third result states that the
finite-size flow at a fixed cut-off is governed by a standard
beta-function relation, where the functions are simply related to
those for the change in $N$:
\be
\bb{i}(\kappa;N) = y \kapp{i}  + \bt{i}(\kappa;N)
\;.
\ee
Finally, a conformal defect corresponds to a zero of the beta-functions
$\bb{i}(\kappa;N)$. From the leading behaviour \eqref{eq:betat}, the
positions of the zeroes $\kappp{}{*}a(N)$ scale approximately with $N$ as
$N^y$ which we check in one case in section \ref{sec:TCSA-results}.

Given the TCSA Hamiltonian, it is possible to calculate the TCSA
approximation of many different physical quantities, but in this paper
we have examined only one, namely the spectrum of the Hamiltonian,
and we describe in the next section how we use this to identify
the presence of possible RG fixed points, that is conformal defects,
and attempt to identify them.

\subsection{Identification of conformal defects using TCSA}

From the TCSA method, we obtain the spectrum of the Hamiltonian on a
strip. The simplest way 
to identify candidate conformal defects  is to place the defect in the
middle of the strip and use the fact that in this case the partition
function takes the special form \eqref{eq:Zfolded} and the spectrum is
organised into representations of the Virasoro algebra of central
charge $2c$.  Consequently, we examine the TCSA results to find
spectra which take  this special form and propose these as candidates
for conformal defects. 

To be explicit, we consider the model on a strip of width $R$ with
conformal boundary conditions on the two edges and the
defect $D$ in the middle.  In our TCSA calculations we will always
take the boundary condition to be the same on both edges, but for the
moment we shall label them $B$ and $B'$ for generality.
This system can be folded into the folded model of
central charge $2c$ on a strip of width $R/2$ with a factorised
boundary condition $(B,B')$ on one edge and a boundary condition
corresponding to the defect on the other edge. If this defect is
conformal, then the spectrum falls into representations of the
Virasoro algebra of charge $2c$ and
\be
 Z_{(B,B');D_{conformal}} = \sum_h m_h \chi_{h,2c}(q)
\;,
\labl{eq:Zstrip1}
where $q=\exp(-2\pi L/R)$. 
This means that the spectrum falls into sets of levels separated by
integer multiples of $2\pi/R$ with distinctive multiplicities.
If these are absent, then the defect $D$ cannot be conformal.

For the two special cases of a topological defect and a factorised
defect the spectrum takes more particular forms.

If the defect $D$ is  a topological defect 
then the partition function is simply that of the original model on a
strip of width $R$ with two conformal boundary conditions, so that 
\be
   Z_{(B,B');D_{topological}} 
= \sum_{i} m_i  \chi_{i,c}(\sqrt q)
\;.
\labl{eq:Zstrip2}
In this case,  the spectrum falls into sets of levels separated by
integer multiples of $\pi/R$, which is half the spacing of the energy
levels in the general case.

If the defect $D$ is in fact a factorised defect of the form
\eqref{eq:Fform} then the partition function takes the special 
form
\be
   Z_{(B,B');D_{factorised}} 
= \sum_{a,b,i,j} \left(n_{ab} N_{Ba}^i N_{B'b}^j\right)
 \chi_{i,c}(q) \chi_{j,c}(q)
\;.
\labl{eq:Zstrip3}
In this case,  the spectrum falls into sets of levels separated by
integer multiples of $2\pi/R$ as in the general case but with typically
much higher degeneracies than those of a general conformal defect. 

A further check can be obtained by moving  the defect across the strip
to different angles $\theta$. For topological defects the spectrum
should be independent of $\theta$ while for purely factorising defects
$F=\sum_{a,b} n_{ab}\,\vvecc{a}\,\ccevv{b}$ the partition
function is
\be
   Z_{(B,B');D_{factorised}}(q;\eta)
= \sum_{a,b,i,j} \left(n_{ab} N_{Ba}^i N_{B'b}^j\right)
  \chi_{i,c}(q^{\frac1{2\eta}})\chi_{j,c}(q^{\frac1{2(1-\eta)}})\,,
\quad \theta = \eta\,\pi\,,
\labl{eq:Z_factorised}
which lets us identify the coefficients $n_{ab}$ exactly.

\sect{TCSA results}
\label{sec:TCSA-results}

We consider two different systems: firstly the Ising model, as a check
on our method and confirming the results of \cite{Fendley:2009gm}; 
secondly, we consider minimal models with $p$ large for which we take
$p=10$ as a typical example. Throughout this section we use the
dimensionless couplings $\kapp{i}$ as appropriate to the TCSA method.

\subsection{The critical Ising model}\label{sec:Ising}

\begin{figure}[tb]
\subfigure[TCSA results in $M(3,4)$ for $\alpha=3\pi/8$ plotted
against $\log(\kappa)$ for $N=24$ and 762 states. The degeneracies of 
the energy levels at the IR fixed point are grouped according to the
representation of the $c=1$ folded model into which they fall.]
{\scalebox{1.0}{
  \begin{picture}(220,220)
  \put(0,0){\scalebox{0.7}{\includegraphics{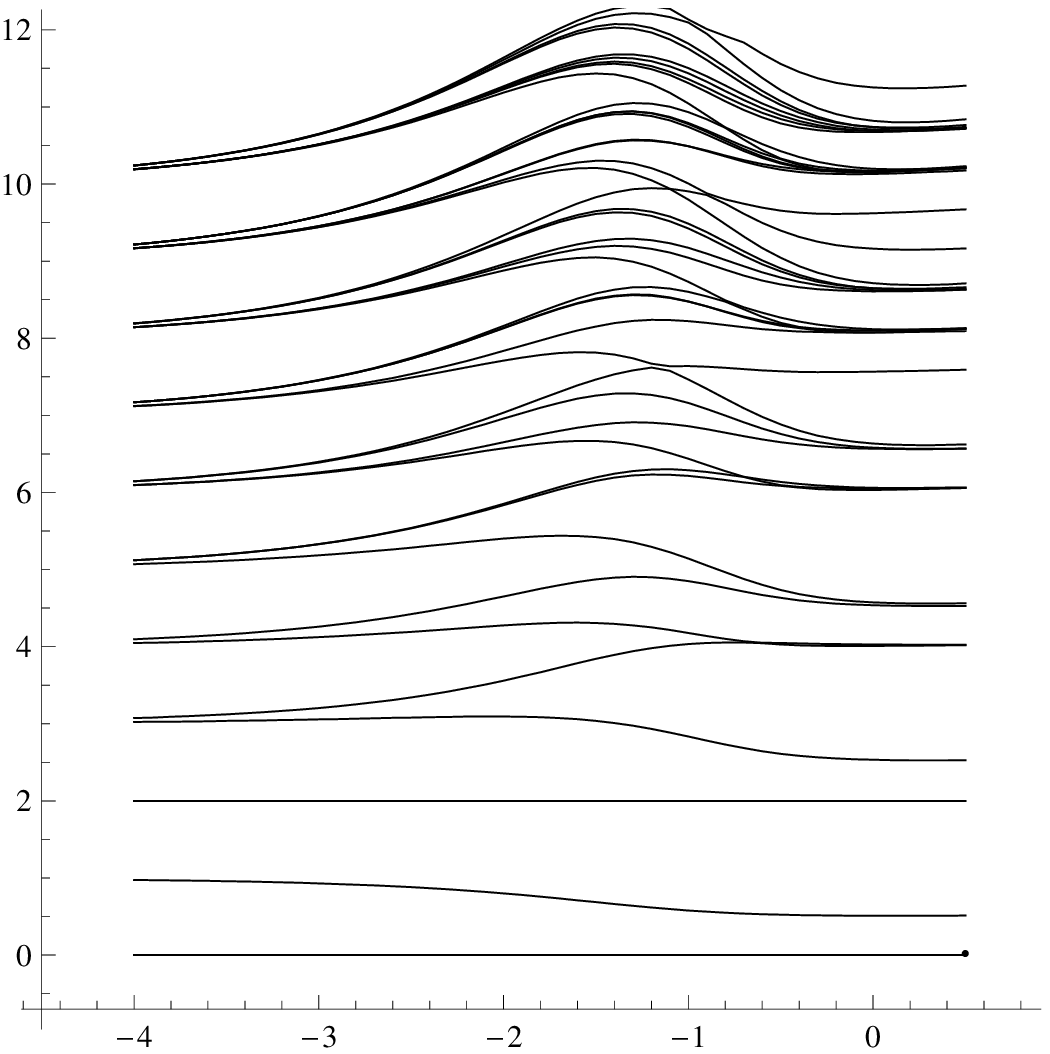}}}
  \put(0,0){
     \setlength{\unitlength}{.8truept}\put(-90,-587){
     \put(111,608)   {\scriptsize 1}
     \put(111,628)   {\scriptsize 1}
     \put(111,648)   {\scriptsize 1}
     \put(111,668)   {\scriptsize 2}
     \put(111,688)   {\scriptsize 2}
     \put(111,708)   {\scriptsize 3}
     \put(111,728)   {\scriptsize 4}
     \put(111,748)   {\scriptsize 5}
     \put(111,768)   {\scriptsize 6}
     \put(111,788)   {\scriptsize 8}
     \put(109,808)   {\scriptsize 10}
     \put(338,608)   {\scriptsize 1} 
     \put(338,648)   {\scriptsize 1}
     \put(338,688)   {\scriptsize 2}
     \put(338,728)   {\scriptsize 3}
     \put(338,768)   {\scriptsize 5}
     \put(338,808)   {\scriptsize 7}
     \put(342,620)   {\scriptsize 1} 
     \put(342,660)   {\scriptsize 1}
     \put(342,700)   {\scriptsize 2}
     \put(342,740)   {\scriptsize 3}
     \put(342,780)   {\scriptsize 5}
     \put(342,820)   {\scriptsize 7}
     \put(346,760)   {\scriptsize 1} 
     \put(346,800)   {\scriptsize 1}
     \put(350,789)   {\scriptsize 1}
     \put(350,830)   {\scriptsize 1}
     }\setlength{\unitlength}{1pt}}
  \end{picture}
}
\label{fig:isingflow}}
\hfill
\hspace{3mm}
\subfigure[The spectrum of $M(3,4)$ at $\log(\kappa)=0.3$ as a function of
$\alpha$ for for $\pi/4 \leq  \alpha \leq 5\pi/4$ with $N=24$ and 762 states.]%
{\scalebox{0.70}{\includegraphics{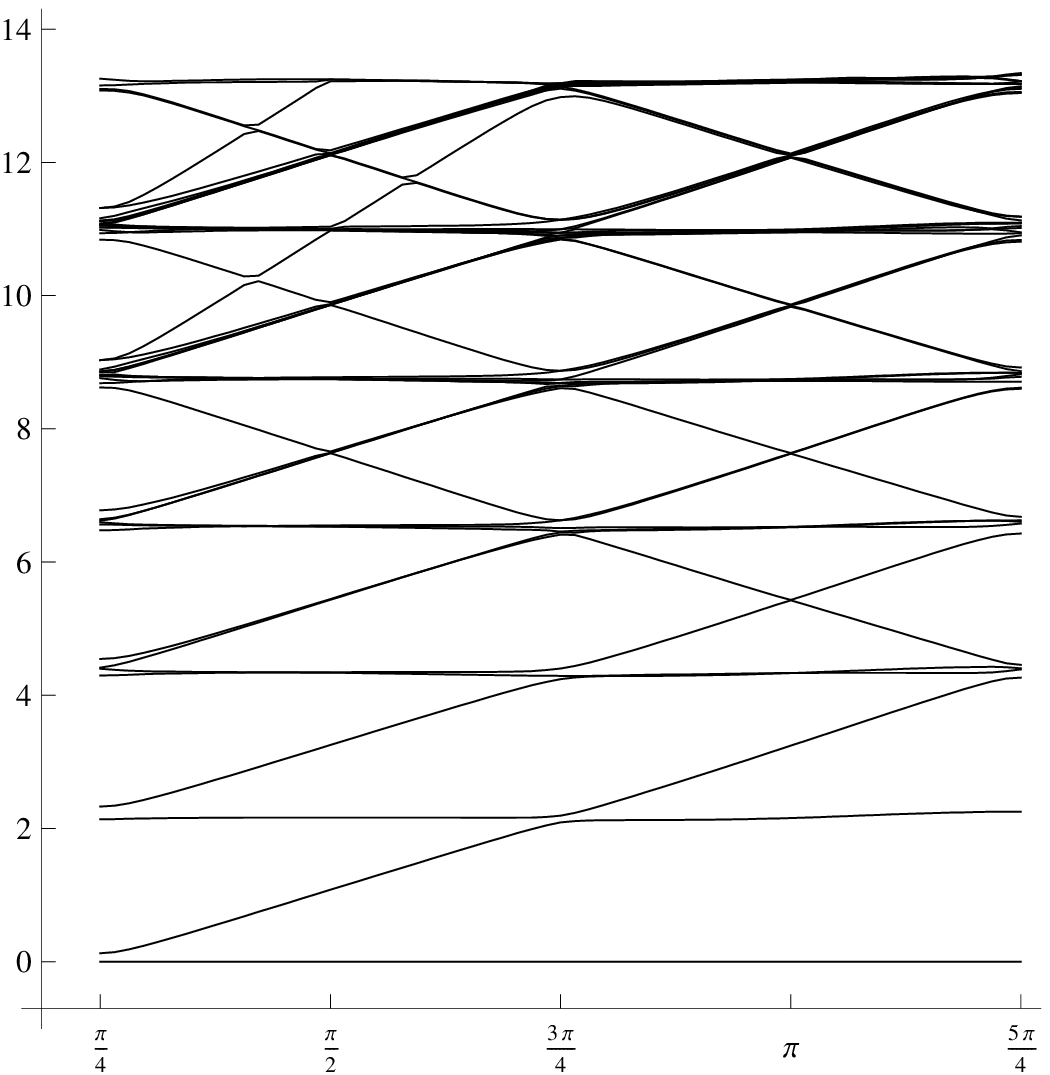}}\label{fig:isingroundpl}}
\caption{IR fixed points in the critical Ising model}
\end{figure}

Using the defect TCSA method we can check the predictions of
\cite{Fendley:2009gm} as well as the accuracy of the TCSA method. 
We recall that \cite{Fendley:2009gm} predict flows of the form
$D_\sigma \rightarrow N(5\pi/4 - \alpha)$ where 
\be
\tan(\alpha)=\kapr/\kapl\,,
\labl{eq:alpha}
As an example, we calculate the TCSA spectrum for
$\alpha=3\pi/8$ and compare it with the partition function at
$\fii_0=5\pi/4-3\pi/8=7\pi/8$ given in \eqref{eq:Z.OA}
\be\begin{array}{ll}\displaystyle
Z_{D(0)_+,D(7\pi/8)}(q;\frac12) &\!\!=
q^{\frac{49}{128}-\frac1{24}}(1+q^{\frac14}+q+q^{\frac54}+2q^2+2q^{\frac94}
\\[.5em]
& \quad +\,3q^3+3q^{\frac{13}4}+q^{\frac{15}4}+5q^4+5q^{\frac{17}4}+q^{\frac92}+q^{\frac{19}4}+
7q^5+\dots )\,.
\end{array}
\labl{eq:Zexample}
In figure \ref{fig:isingflow} the normalised energy differences
$2 (E_i-E_0)/(E_2-E_0)$ 
are shown against the logarithm of the
dimensionless coupling strength $\kappa$ of the perturbation. 
With this normalisation, the indicator of an IR fixed point is the
existence of a state with normalised gap 4. The only such point
visible is that with $\log(\kappa)\sim 0.3$. At this point, the
energy levels visible have rearranged themselves into 
four distinct sets and both this arrangement and the values of
the energy gaps agree with the partition function 
\erf{eq:Zexample}. 
We conclude that the endpoint of the flow starting at
the angle $\alpha=3\pi/8$ in the $(\kapl,\kapr)$ plane is indeed the
conformal defect $D(5\pi/4-3\pi/8)=D(7\pi/8)$.

The IR fixed point is located at a finite value of the bare coupling
constant and at approximately the same
distance from the origin in every direction. This means we can view
all the IR fixed points together by plotting the energy differences
$\frac{r}\pi(E_i-E_0)$ against the angle $\alpha$ at this fixed value
of $\kappa$. Expanding the partition function \erf{eq:Z.OA} gives
\be 
Z_{D(0)_+,D(\fii_0)}(q;\frac12) =
q^{\frac{x^2}2-\frac1{24}}(1+q^{2-2x}+q+q^{3-2x}+2q^2+2q^{4-2x}+q^{2+2x}+\dots)\,,
\ee 
where $x=\fii_0/\pi$. This means that the energy gaps are straight
lines as functions of $\fii_0$ or $\alpha$ and this is exactly what
can be seen in the TCSA plot (figure \ref{fig:isingroundpl}). 
The fact that these lines are straight also implies that
$\kapp{1}/\kapp{2}$ is a constant along the finite-size scaling
flow. This is not expected to be the case in general, but the extra
symmetry in the Ising model allows this to happen.

This result
provides a very strong indication that our identification of the RG
flow fixed points in terms of Dirichlet-type defects is correct,
confirming both the calculations of \cite{Fendley:2009gm} and the usefulness of
the TCSA method.

\subsection{Minimal models with $p>3$}\label{sec:p>3}

For the minimal models $M(p,p+1)$ with $p>3$ the flow picture is
different from that for the critical Ising model, as is to be expected
from the RG analysis. The proposed
landscape of flows can be seen in figure \ref{fig:flow-p>3}. In the
following we illustrate the exploration of this picture using the
$M(10,11)$ model with TCSA. 

\subsubsection{Chiral perturbations}

\begin{figure}[tb]
\subfigure[positive direction: $(1,2)\to(2,1)\dashleftarrow(1,3)$]{\scalebox{0.7}{\includegraphics{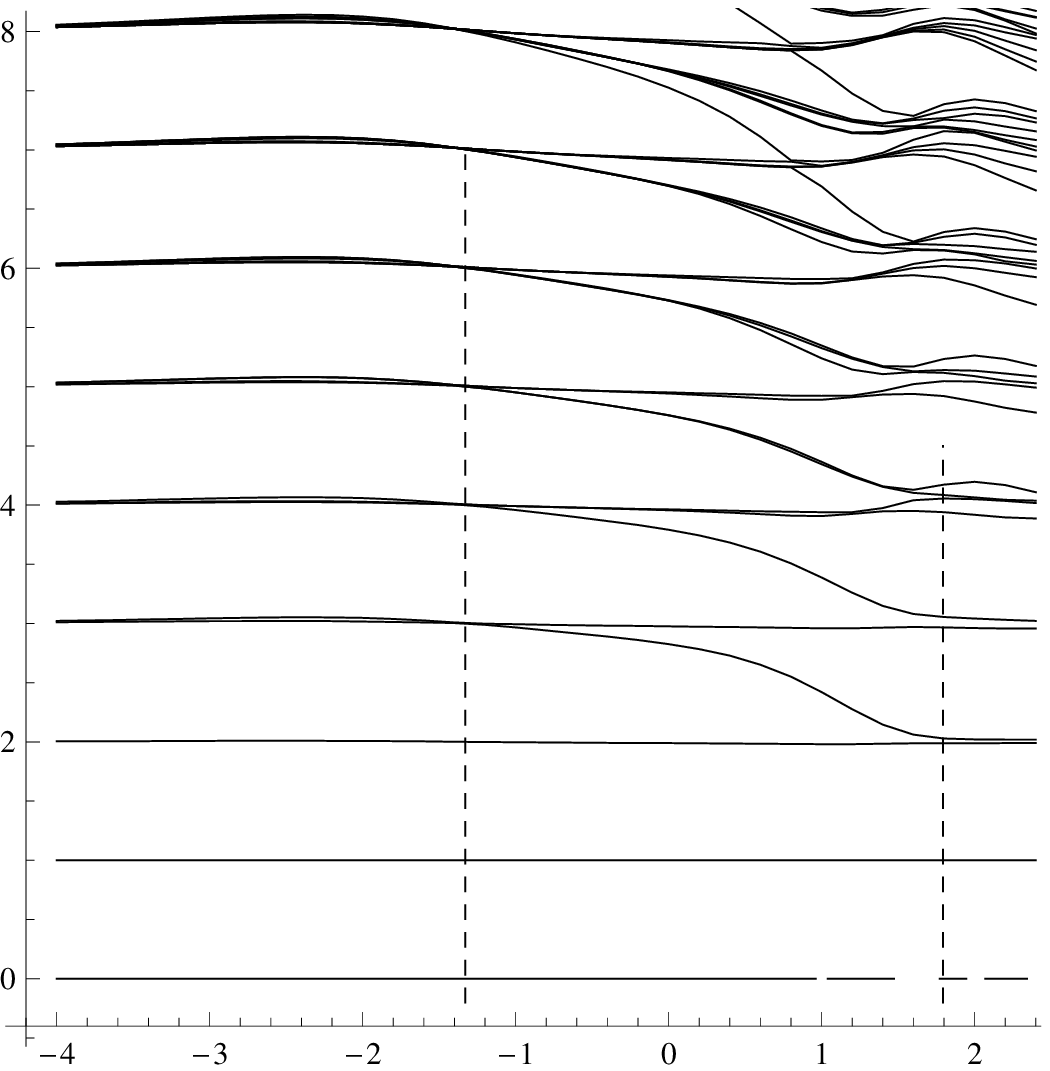}}}
\hfill
\subfigure[negative direction:$(1,2)\to(1,1)$]{\scalebox{0.7}{\includegraphics{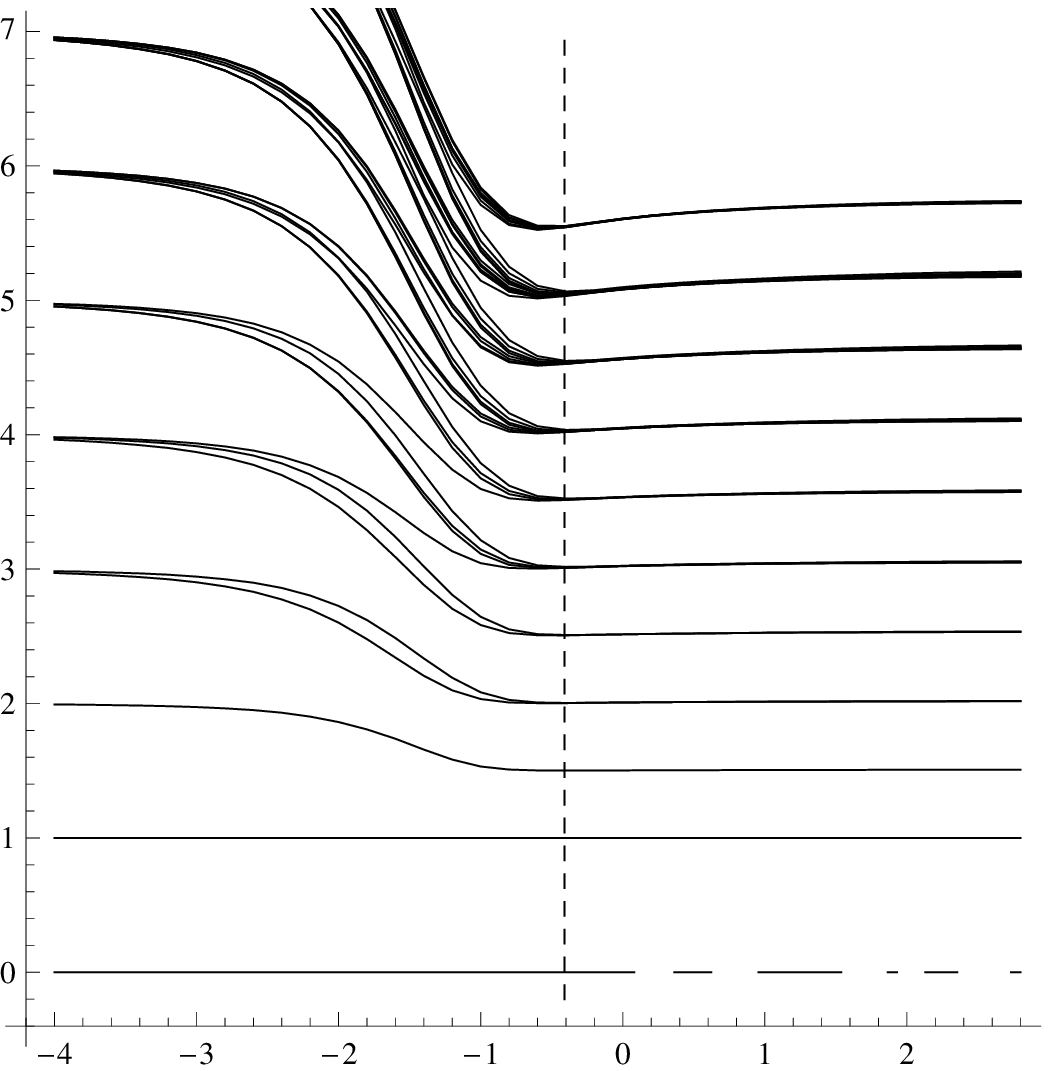}}}
\caption{$M(10,11)$: chiral flows. We show the normalised energy
  levels plotted against $\log(\kappa)$ with $N=22$, 1794 states.}
\label{fig:10.11chir}
\end{figure}

The simplest cases are the purely chiral perturbations. The spectra
for these models are both formally and numerically identical to the 
corresponding boundary perturbations 
as is explained in section \ref{sec:chirdefpert}.
From section \ref{ss:chiral} we see that the IR
fixed 
points in the positive and
negative directions are the $(2,1)$ and the $(1,1)$ defects,
respectively. These  
defect flows are shown in figure 
\ref{fig:10.11chir} where the normalised
energy differences are 
plotted against the logarithm of the coupling strength. In the
positive direction the cut-off effects drive the flow past the first
fixed point to a second one, realising the reversed version of the
$(1,3)\to(2,1)$ flow. Since the spectra are numerically identical to
those in the boundary case, this effect is identical to that observed for
boundary flows in \cite{Feverati:2006ni}. 

In the perturbative direction we can estimate the location of the TCSA
fixed point as the value of $\kappa$ for which the fourth and fifth
energy levels cross and this depends on $N$ as given in table
\ref{tab:n-dependence1}. As can be seen it is very close to the
approximate value predicted by the analysis in \cite{Feverati:2006ni}
which gives $\kappp{}{*} \approx 0.1526 N^y$, once the different
normalisation of the TCSA perturbing field is taken into account.

\begin{table}
$$
{\renewcommand{\arraystretch}{1.4}
\begin{array}{c|ccccccc}
N & 10 & 12 & 14 & 16 & 18 & 20 & 22  \\
\hline
\log({\kappp{}{*}}_{\mathrm{measured}}) &
%-1.513  & -1.465  & -1.428  & -1.397  & -1.371  & -1.348  &\\
-1.51  & -1.46  & -1.43  & -1.40  & -1.37  & -1.35  & -1.33\\
\log({\kappp{}{*}}_{\mathrm{theory}}) &
%-1.461  & -1.428  & -1.400  & -1.376  & -1.354  & -1.335 &\\
-1.46  & -1.43  & -1.40  & -1.38  & -1.35  & -1.34 & -1.32\\

\end{array}
}$$
\caption{The approximate position of the perturbative fixed point in
  $M(10,11)$ in the purely chiral direction as a function of TCSA
  cut-off $N$
}
\label{tab:n-dependence1}
\end{table}

\subsubsection{The third quadrant: $\kapl<0$, $\kapr<0$}
\label{sec:mm_dir}

Next we turn to the third quadrant of the plane of the couplings in
figure \ref{fig:flow-p>3}.
We have searched thoroughly for fixed points in the domain of
convergence of TCSA and found  a single fixed
point $F$, which is in the diagonal direction $\kapl=\kapr<0$. 

\begin{figure}[t!]
\begin{centering}
\scalebox{0.675}{  \begin{picture}(300,300)
  \put(0,0){\scalebox{1}{\includegraphics{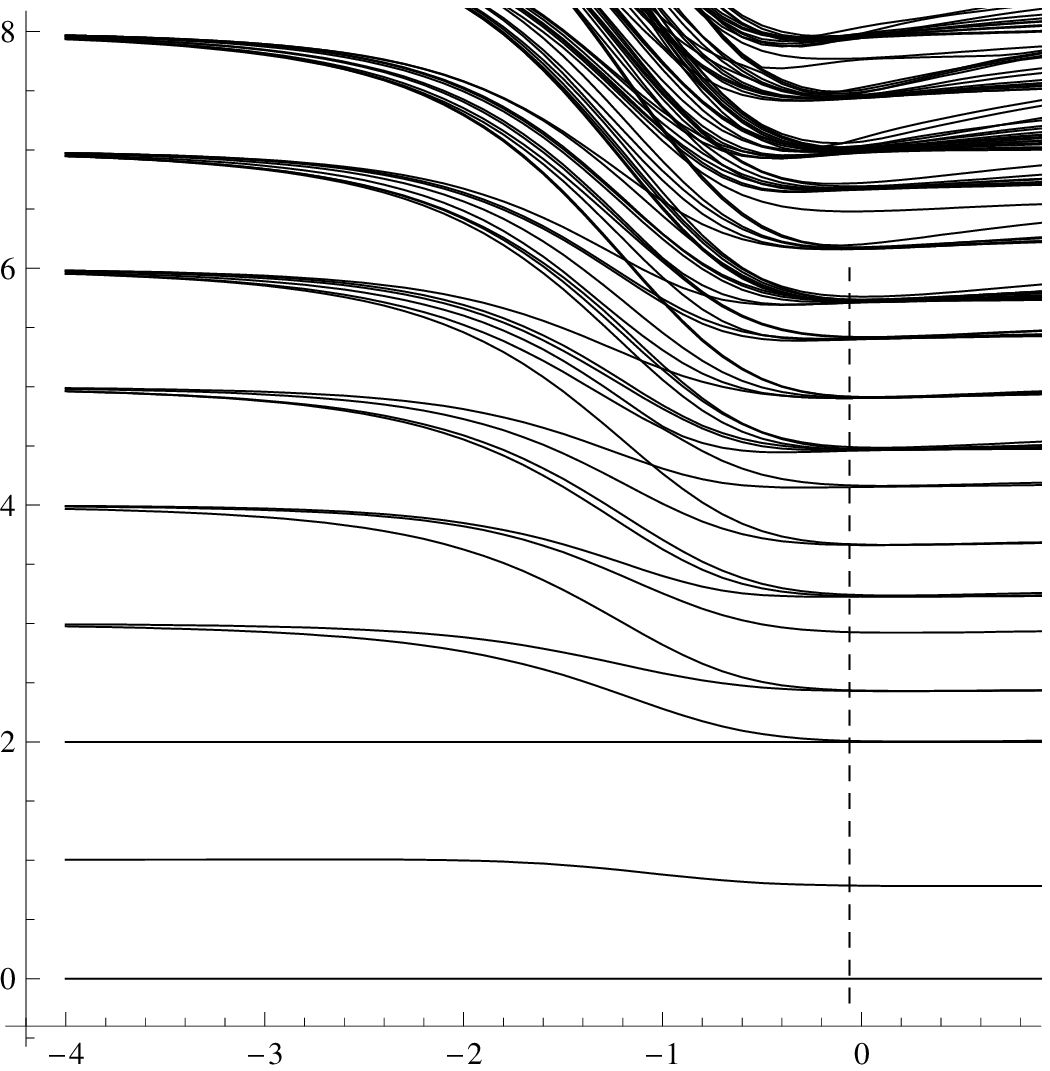}}}
  \end{picture}}
\\
\subfigure[Even sector]
{
  \begin{picture}(150,150)
  \put(0,0){\scalebox{0.45}{\includegraphics{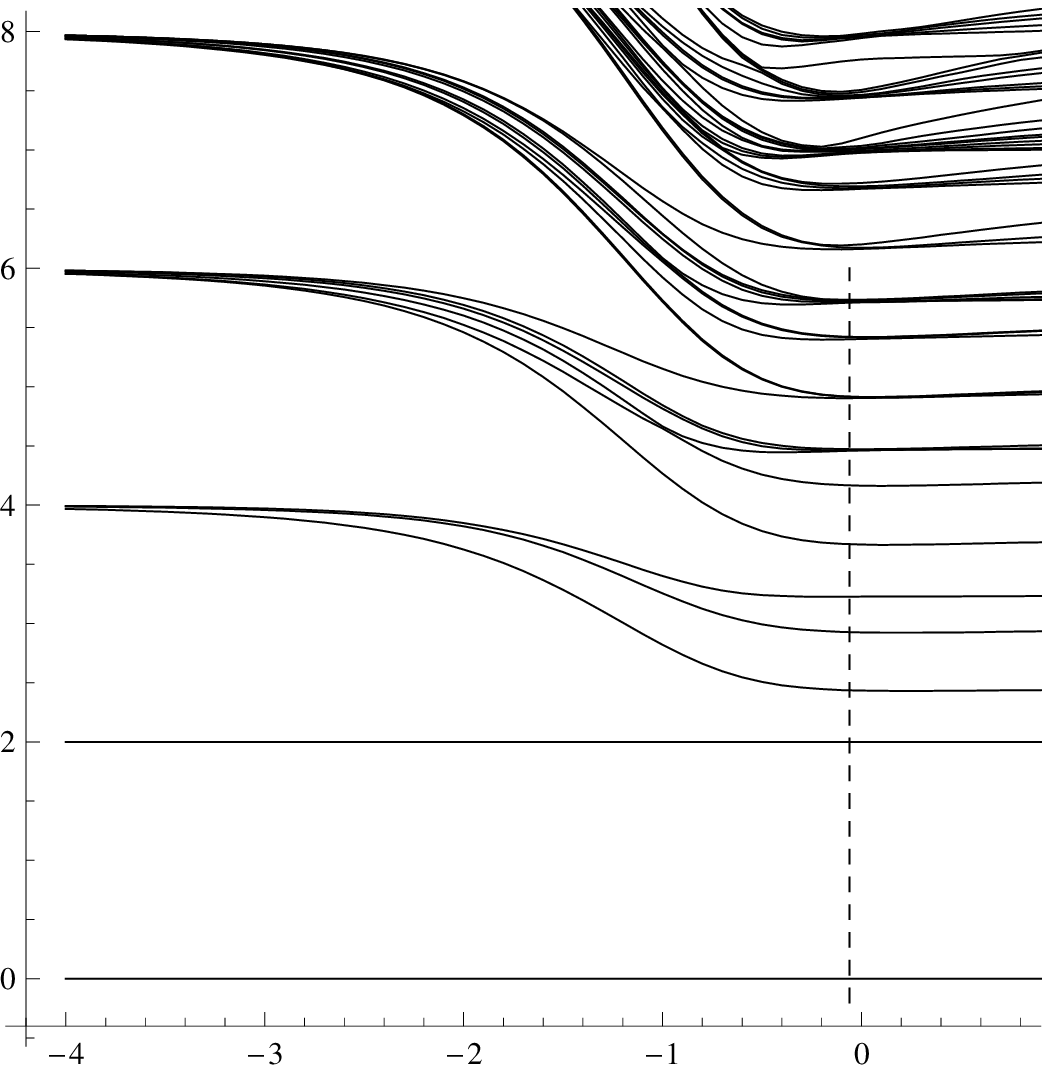}}}
  \end{picture}
}
\subfigure[Odd sector]
{
  \begin{picture}(150,150)
  \put(0,0){\scalebox{0.45}{\includegraphics{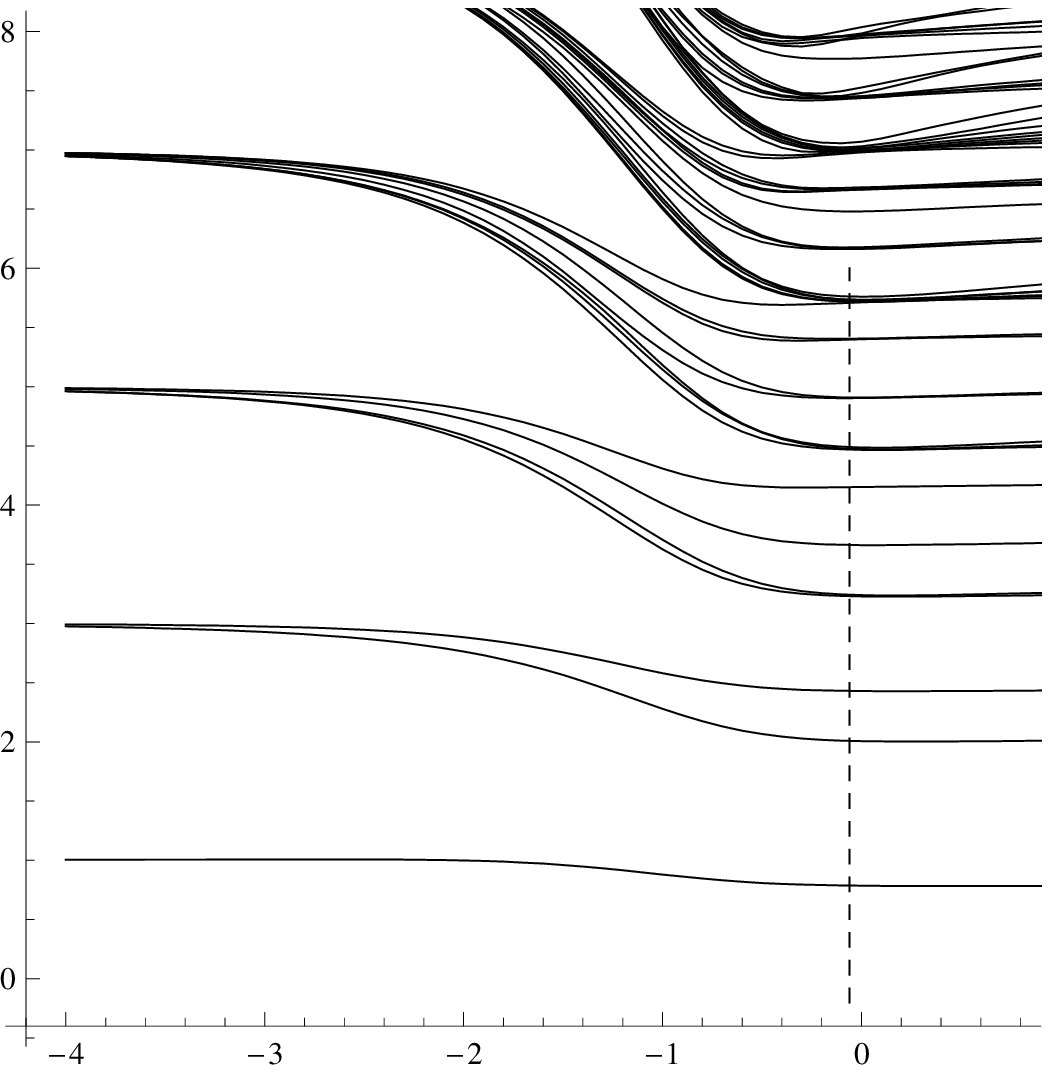}}}
  \end{picture}
}
\\
\end{centering}

\caption{$M(10,11)$: diagonal flow in the negative direction. 
In the main
  figure, the scaled energy gaps for both sectors are shown, the odd
  sector with TCSA cut 19 and the even sector 
  given by the combination of cuts 18 and 20 as detailed in the text.
  The small sub-figures show the even and odd sectors separately.
}
\label{fig:10.11mmall}  
\end{figure}

As discussed in section \ref{ssec:TCSA-Ham}, the TCSA Hamiltonian for
$\kapl{=}\kapr$ is
block diagonal and the flows for the two sectors are shown in figure 
\ref{fig:10.11mmall}. 
We find that for large $\kappa$ the spectrum approaches an IR fixed
point for which the eigenstates in each sector separately fit into
representations of the folded model. 

Combining the two sectors is a
non-trivial task, because they have different effective cut-offs.
This means that they have different values for the non-universal
boundary energy and different re-scalings of the couplings $\kappa$ and
$r$ so that the two sectors cannot be directly compared.
The usual perturbative renormalisation
based on minimal subtraction or analytic continuation cannot be
implemented within the TCSA framework, but it is possible to define a
prescription for the relative shift of the sectors. Going from a TCSA
cut $N=2n$ to $N=2n+1$ the even energy levels are not affected,
because the new states in the Hilbert space belong to the odd
sector. Similarly, raising the cut from $N=2n+1$ to $N=2n+2$ the odd
levels do not change. One can define, say, the values of the even
eigenvalues at the odd cut $N=2n+1$ as
\be
e_i^\text{even}(2n+1)=\frac12\left(e_i^\text{even}(2n)+e_i^\text{even}(2n+2)\right)\,.
\labl{eq:shiftrule}
The justification of this prescription is provided by its
application to the fixed point $F$ where it yields the spectrum shown
in figure \ref{fig:10.11mmall}. It corresponds to
a partition function of the form \eqref{eq:Z_factorised} with
$\eta=1/2$, namely
\be
Z_F(q;\frac12)=\sum_{r=1}^9\chi_{r,1}(q)^2\,,
\ee
although we could not identify all the components as some of them
enter only at high level.

This implies that 

\medskip
{\it the IR fixed point in the negative diagonal direction is the
factorising defect \[F=\sum^{p-1}_{r=1}|\vec{r,1}\rangle\langle\cev{r,1}|\,.\]}
We found the same result for other minimal models and we checked also
that the dependence of the spectrum on the position of the defect is
as it is expected and given by \eqref{eq:Z_factorised}. Note that the
defect $F$ is of the form of
\eqref{eq:C-fact-rest} with $s=s'=1$.

\newpage
\subsubsection{The second and fourth quadrants: $\kapl$ and $\kapr$
  of opposite signs}

\begin{figure}[tb]
\subfigure[round plot]{\scalebox{0.7}{\includegraphics{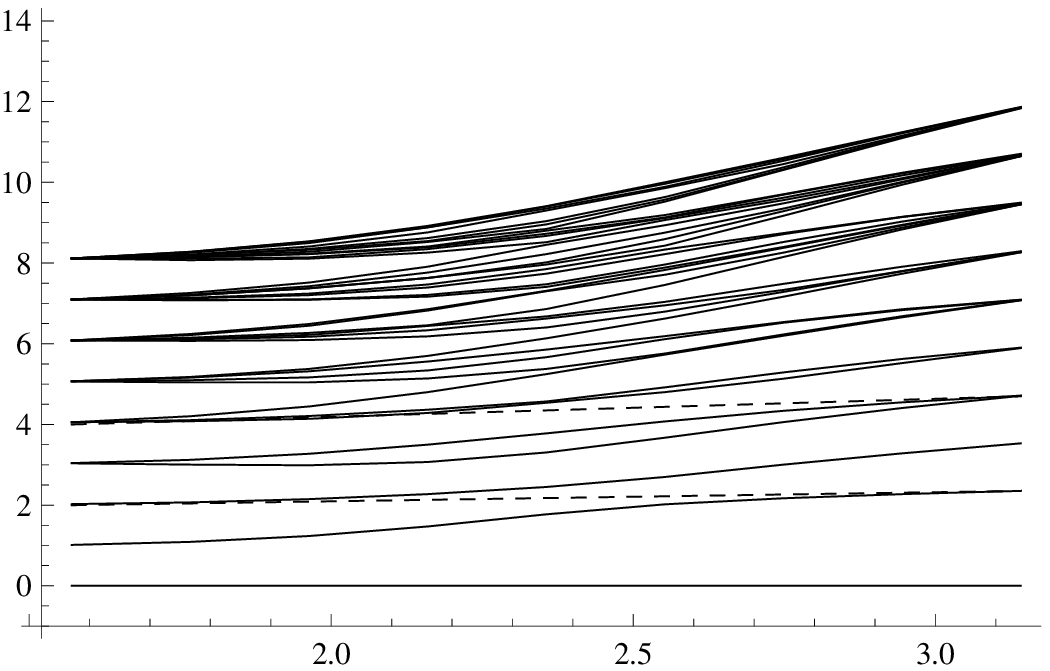}}%
\label{fig:10.11round_m}}
\hfill
\subfigure[$(2,1)-\phi_{13,33}$ flow]{\scalebox{0.7}{\includegraphics{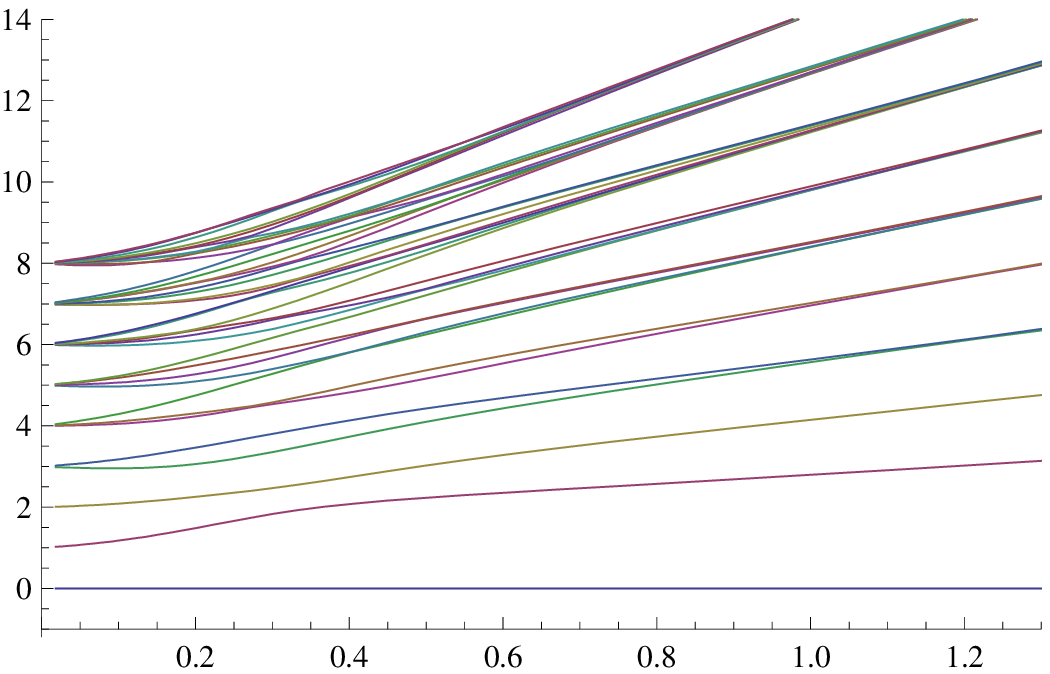}}%
\label{fig:10.11_21m}}
\caption{$M(10,11)$: comparison between the $(2,1)-\phi_{13,33}$ flow
  and the round plot. The absence of states of scaled energy 2 and 4
  means that there is no indication of a conformal fixed point between 
the points $\alpha=\pi/2$ and $\alpha=\pi$. The different slopes of
the lines is a result of the different renormalisations of the
function $r(\kappa)$ in the two systems.}  
\end{figure}

For different signs of the couplings $\kapl$ and $\kapr$ we found 
strong indications of a fixed point far from the perturbative region
which appears to be a deformation of the $(ff)$ fixed point in the
Ising model;  
we call this conformal defect $(ff)'$. However, since this is far from
the perturbative 
region, there is no necessity that this be reachable by a finite-size
scaling flow
from the defect $D$, and indeed we believe that all RG flows starting
at $D$ in this quadrant end up at the identity defect $I$. 
This is the result predicted by the simple model in section
\ref{sec:sum} and to test it we explored the
quadrant by making ``round plots'' similar to the Ising model, that is
plots of the spectrum as functions of 
   $\alpha$ at various fixed $\kappa$ for the Hamiltonian \eqref{eq:two-k-ham}.
In figure 
\ref{fig:10.11round_m} the weight
differences are plotted against the angle $\alpha$, defined in the
same way as for the Ising model in 
\eqref{eq:alpha}, with $\log(\kappa)$ interpolating between the
approximate fixed point values $\log(\kappa^*){=}-1.4$ at
$\theta=\pi/2$ and $\log(\kappa^*)=-0.4$ at $\theta=\pi$. The plot
does not feature any fixed points but looks 
like a flow interpolating the chiral fixed points $(2,1)\to(1,1)$
located on the vertical and horizontal axes. This picture can be
compared with an actual flow starting from the defect $(2,1)$. As 
can be seen in table \eqref{tab:weights}, in first order perturbation
theory the field $\phi$ transforms into the relevant field
$\phi_{(13)(33)}$ along the flow $(1,2)\to(2,1)$ triggered by
$\bar\phi$. It is possible to study the flow starting from the defect
$(2,1)$ generated by the non-chiral perturbation $\phi_{(13)(33)}$ in
TCSA; some details of the method are given in Appendix
\ref{sec:TCSA_app}. For negative values of the perturbation the flow
is shown in figure \ref{fig:10.11_21m}. It looks
very much the same as the round plot, as shown 
in the schematic figure \ref{fig:roundplot}(a).

This  tells us that:

\medskip
{\it All RG flows starting from the defect $D$ with $\kapl$ and
  $\kapr$ of opposite signs end at the identity defect $I$, and the
  boundary of this space of flows is an RG flow from $D_{(2,1)}$ to $I$.}

\subsubsection{The first quadrant $\kapl>0$, $\kapr>0$}

\begin{figure}[tb]
\begin{centering}
\scalebox{0.75}{
  \begin{picture}(300,300)
  \put(0,0){\scalebox{1}{\includegraphics{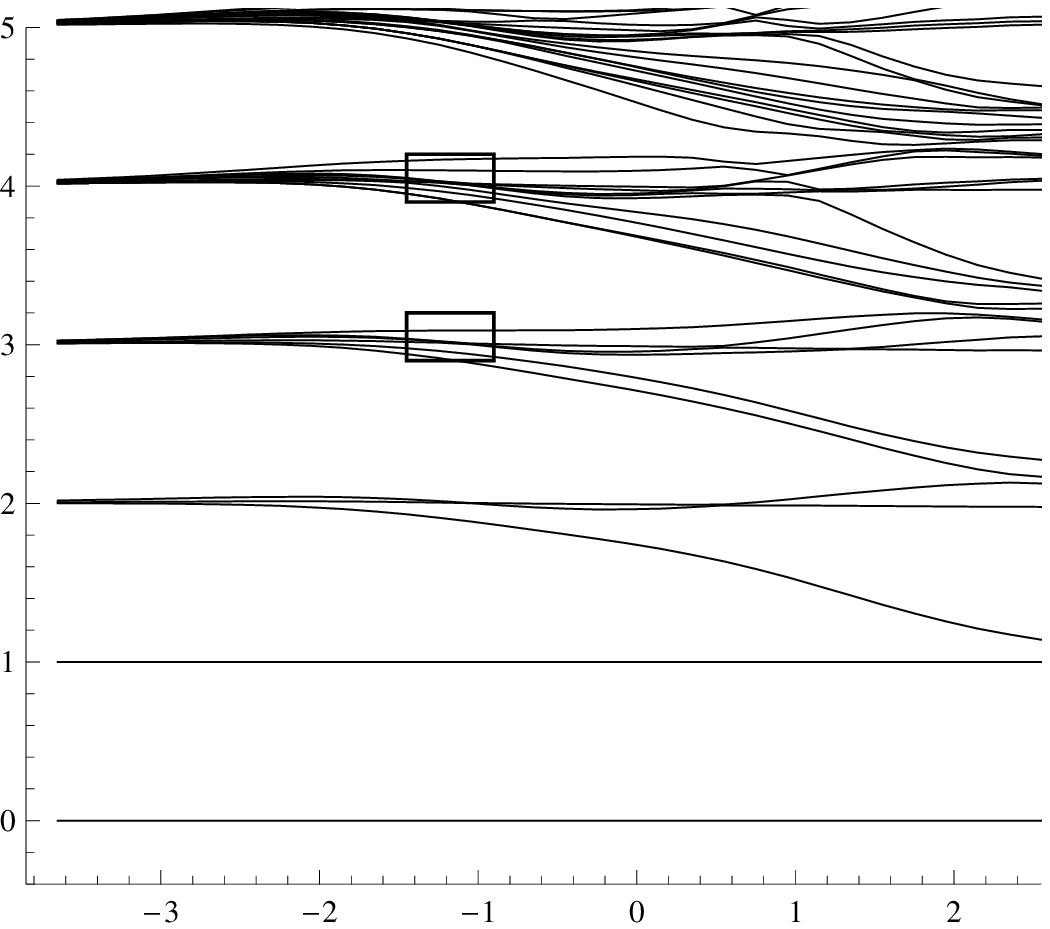}}}
  \put(0,0){
     \setlength{\unitlength}{1pt}{
     \put(130,35)   {$1^a$}
     \put(130,82)   {$1^a$}
     \put(130,106)   {$1^b$}
     \put(130,129)   {$2^a$}
     }}
  \end{picture}}
\\
\subfigure[4th group of lines]
{
\scalebox{0.75}{
  \begin{picture}(150,150)
  \put(0,0){\scalebox{0.5}{\includegraphics{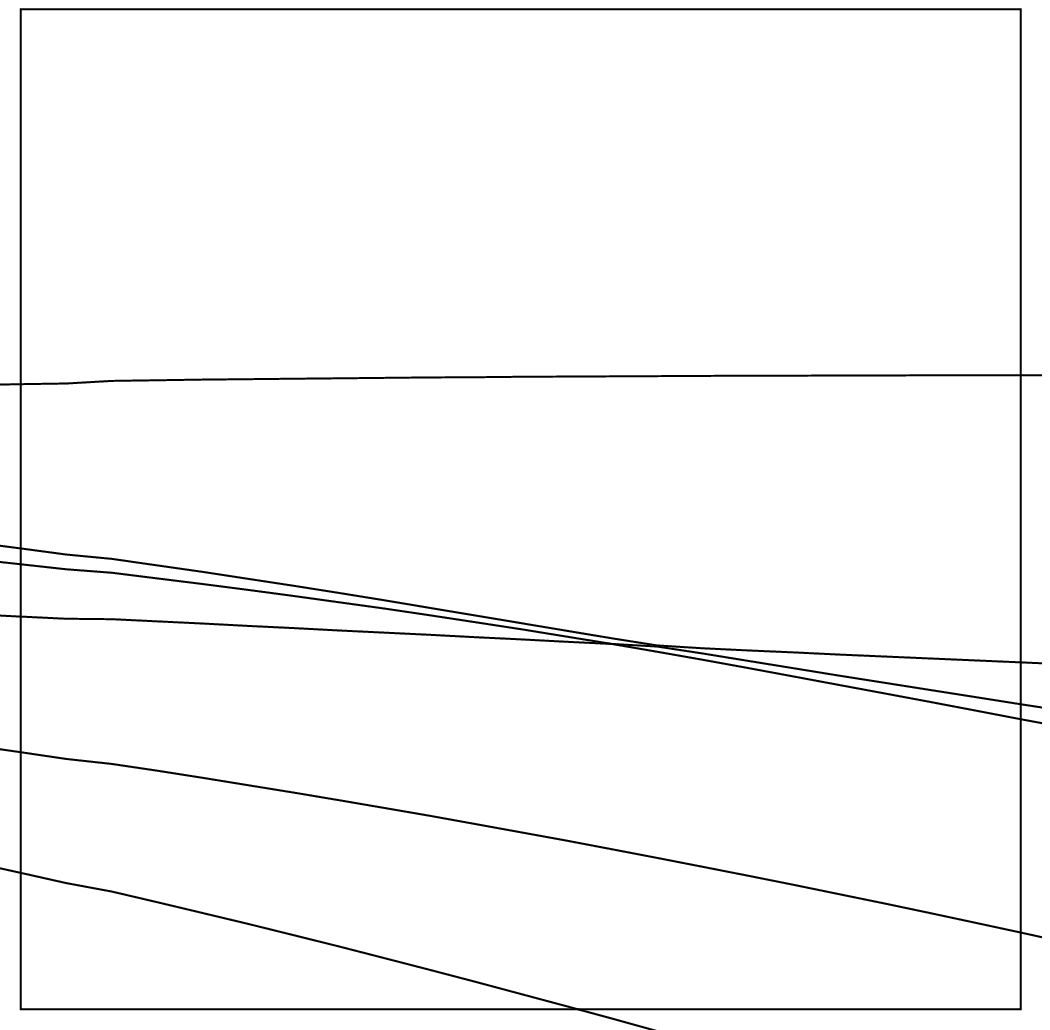}}}
  \put(0,0){
     \setlength{\unitlength}{1pt}{
     \put(79,60)   {$3^a$}
     \put(79,12)   {$1^b$}
     \put(79,35)   {$1^c$}
     \put(79,100)   {$1^d$}
     }}
  \end{picture}
}
}
\subfigure[5th group of lines]
{
 \scalebox{0.75}{
 \begin{picture}(150,150)
  \put(0,0){\scalebox{0.5}{\includegraphics{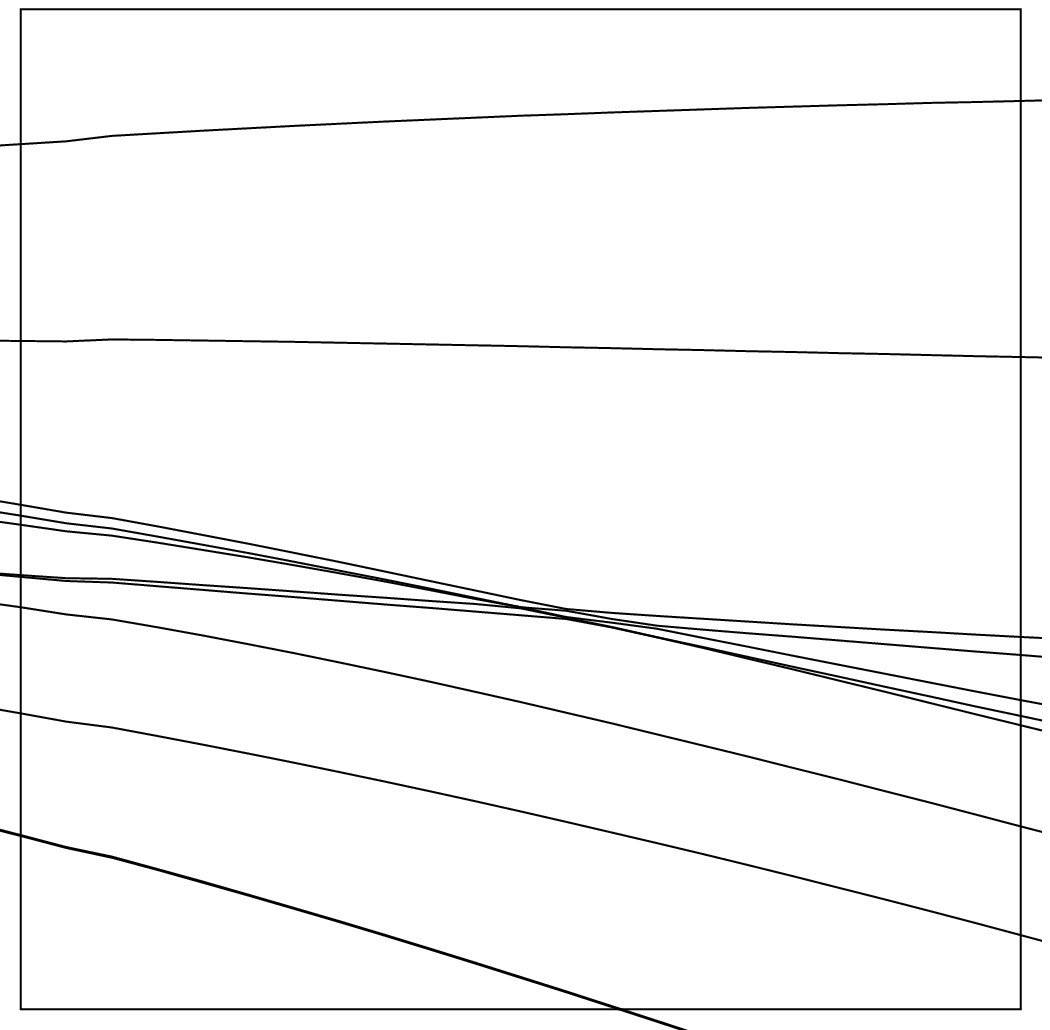}}}
  \put(0,0){
     \setlength{\unitlength}{1pt}{
     \put(79,68)   {$5^a$}
     \put(79,12)   {$2^b$}
     \put(79,35)   {$1^c$}
     \put(79,105)   {$1^d$}
     \put(79,51)   {$1^e$}
     \put(79,135)   {$1^f$}
     }}
  \end{picture}
}}
\\
\end{centering}

\caption{$M(10,11)$: diagonal flow in the positive direction, even
  sector, scaled energy gaps vs $\log(\kappa)$ for a strip of width
  $\pi$ truncated at level 20; the boxed regions are shown enlarged in
  the two smaller figures. The energy levels at the fixed point can
  grouped into representations of the folded model and are labelled
  $m^l$ where $m$ is the degeneracy and $l$ is the label of the representation.
}
\label{fig:10.11pp_even}
\end{figure}

Based on perturbation theory we expect a non-topological fixed point
$C$ in the diagonal direction $\kapl=\kapr>0$. We can apply the tools
used in the previous cases for exploring the flows in the first
quadrant. In figure \ref{fig:10.11pp_even}
the normalised energy differences are plotted
against the logarithm of the coupling strength in the positive
diagonal direction for the even sector.
We find multiple degeneracies at $\log(\kappa)\sim -1.2$ indicating
the presence of a conformal fixed point. 
One feature of this fixed point characteristic of a
perturbative fixed point is that the energy levels stay close to their
UV values, so that lines appear to split apart initially then to
rejoin and intersect with multiple degeneracies at $\kappa^*$.

As shown in the magnified plots of the regions around
the fixed point, the states approximately
group into representations of the folded model with the correct
multiplicities.
An analogous pattern of line crossings is also seen in the odd sector.

\begin{figure}[tb]
\subfigure[The conjectured exact flow
  from $D_{(2,1)}$ to $I=D_{(1,1)}$ and the set of flows starting from
  $D_{(1,2)}$ for various values of $\alpha$ in the second quadrant.]{
$$  
  \begin{picture}(190,160)
  \put(30,0){\scalebox{0.80}{\includegraphics{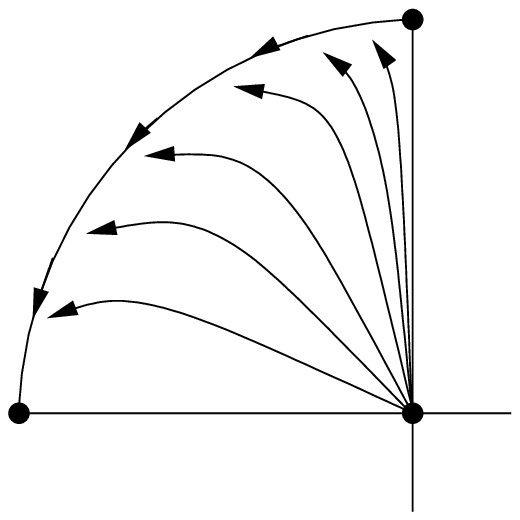}}}
  \put(30,0){
     \setlength{\unitlength}{.80pt}\put(-90,-587){
     \put(88,600)   {$I$}
     \put(212,720)   {$D_{(2,1)}$}
     \put(212,600)   {$D_{(1,2)}$}
     \put(70,724)  {$D_{(2,1)}-\phi_{13,33}$}
     }\setlength{\unitlength}{1pt}}
  \end{picture}
$$
}
\hfill
\subfigure[The conjectured exact flow
  from $D_{(2,1)}$ to $C$ and the set of flows starting from
  $D_{(1,2)}$ for various values of $\alpha$ in the first quadrant.]{
$$  
  \begin{picture}(190,160)
  \put(20,0){\scalebox{0.80}{\includegraphics{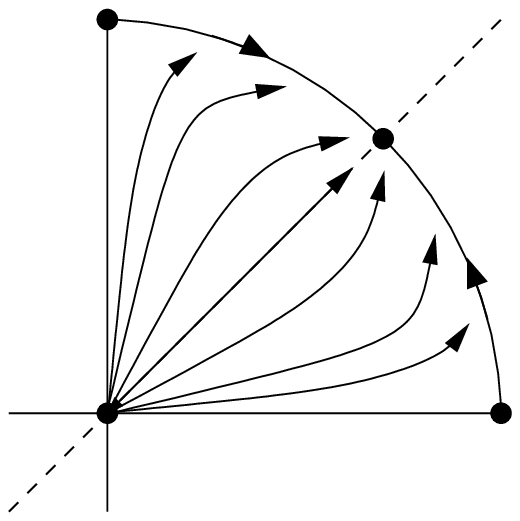}}}
  \put(20,0){
     \setlength{\unitlength}{.80pt}\put(-90,-587){
     \put(170,724)   {$D_{(2,1)}+\phi_{13,33}$}
     \put(235,640)   {$D_{(2,1)}+\phi_{33,13}$}
     \put(207,695)   {$C$}
     \put(83,720)   {$D_{(2,1)}$}
     \put(212,600)   {$D_{(2,1)}$}
     \put(122,600)   {$D_{(1,2)}$}
     }\setlength{\unitlength}{1pt}}
  \end{picture}
$$
}
\caption{The conjectured flows in the first and second quadrants}
\label{fig:roundplot}
\end{figure}

From the perturbative picture we also expect all flows starting with
$\kapl>0, \kapr>0$ to flow to $C$ and the boundary of this space of
flows to be be a pair of flows starting from the two $D_{(2,1)}$
defects generated by $\phi_{(13;33)}$ for $\alpha>\pi/4$ and by
$\phi_{(33;13)}$ for $\alpha<\pi/4$. We have again checked this by
comparing ``roundplots'' in the two-parameter space at fixed $\kappa$
and varying $\alpha$ with the one-parameter TCSA flows
$D_{(2,1)}+\phi_{(13,33)}$ and $D_{(2,1)}+\phi_{(33,13)}$, finding
again qualitative agreement and support for the picture shown in
figure \ref{fig:roundplot}(b).

As the level $p$ of the model $M(p,p+1)$ is reduced, the TCSA
results become less convincing as the new conformal fixed point moves
further away from the UV fixed point. The last model for which there
is convincing evidence of the new defect $C$ is $M(5,6)$; in model
$M(4,5)$ the triple line intersection seen in figure
\ref{fig:10.11pp_even}(a) is no longer seen in TCSA at the trunctation
levels we have been able to achieve. It is possible that increasing
the truncation level would show signs of a conformal fixed point, or
the position of the conformal fixed point may have moved a large
distance; at the moment we cannot tell.

\sect{Conclusion}\label{sec:sum}

\begin{figure}[tb]
\subfigure[The finite-size
  scaling flows in the Ising model showing the ring of conformal
  defects as a function of $\alpha$ with the identification of the
  factorised and topological defects.]{%
$$  
  \begin{picture}(200,100)
  \put(50,0){\scalebox{0.50}{\includegraphics{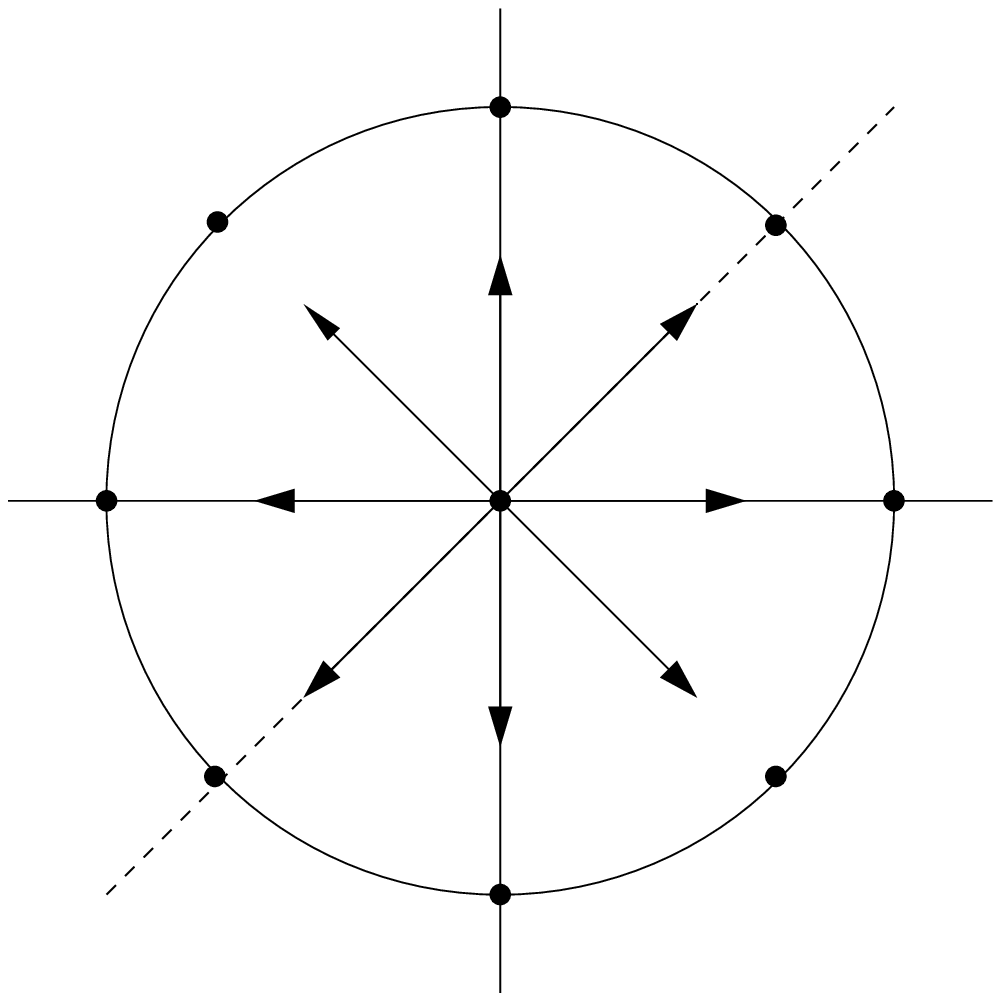}}}
  \put(50,0){
     \setlength{\unitlength}{.50pt}\put(-90,-587){
     \put(100,740)   {$I$}
     \put(239,600)   {$I$}
     \put(250,740)   {$\sigma = D$}
     \put(240,855)   {$\epsilon = D'$}
     \put(355,745)   {$\epsilon = D'$}
     \put(325,810)   {${\scriptstyle (+-)\cup(-+)} = C$}
     \put(5,640)    {${\scriptstyle (++)\cup(--)} = F$}
     \put(100,810)   {$(ff) $}
     \put(320,640)   {$(ff) $}
     }\setlength{\unitlength}{1pt}}
  \end{picture}
$$
}
\hfill
\subfigure[The conjectured pattern of flows out of the\newline defect
$D_{(1,2)}$ showing the 6 IR fixed points for $M(p,p+1)$ with $p$ large]{%
$$  
  \begin{picture}(200,180)
  \put(20,0){\scalebox{0.60}{\includegraphics{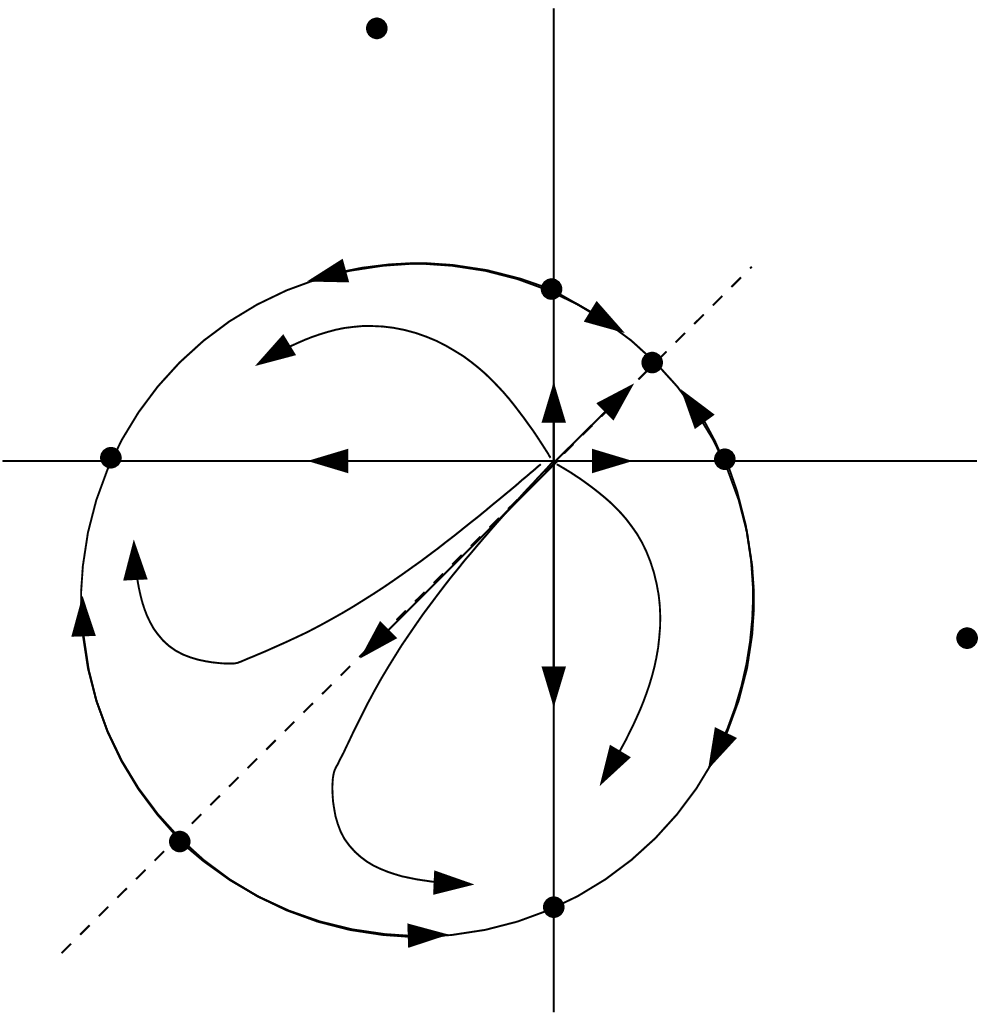}}}
  \put(20,0){
     \setlength{\unitlength}{.60pt}\put(-90,-587){
     \put(105,760)   {$I$}
     \put(255,600)   {$I$}
     \put(255,810)   {$D'$}
     \put(300,760)   {$D'$}
     \put(210,870)   {$(ff)'$}
     \put(380,700)   {$(ff)'$}
     \put(275,790)   {$C$}
     \put(115,632)   {$F$}
     }\setlength{\unitlength}{1pt}}
  \end{picture}
$$}
\caption{Finite-size scaling flows}
\label{fig:finalflows}
\end{figure}

For $p\gg 3$, the perturbative analysis carried out in section
\ref{sec:pert} and the numerical work in section \ref{sec:TCSA-results}
suggests the pattern of flows shown in figure \ref{fig:finalflows}(b). 
We conclude that
for $p\gg 3$ there is at least one IR fixed point which is a conformal defect
that is neither topological nor just a sum over conformal boundary
conditions, namely the one in the $\kapr=\kapl>0$
direction. 

The Ising model $M(3,4)$ behaves differently form the other minimal models
and is treated separately in section \ref{sec:Ising}, with the pattern
of flows \ref{fig:finalflows}(a). 
For $p=3$, the fixed points that have been identified for large $p$
all still remain, but are now just particular points in a continuum.

For $p$ small but greater than $3$, the situation is not so
clear-cut. We are still not sure exactly how the fixed-points approach
the $c=1/2$ continuum, which in part is due to the numerical
inaccuracies of the TCSA method.

The one-loop beta functions as discussed in section \ref{sec:pert}
predict the perturbative fixed points that we find with the TCSA
analysis. It is tempting to try to extend this analysis to higher loop
order to see if they confirm the RG flow structure we suggest in this
paper. As a starting point, we consider the Ising model and then the
changes that are present in the higher minimal models.  Given the
symmetries of the Ising model, the absence of three-point couplings
and the continuum of fixed points, the RG flows away from the $D$
defect can be modelled by the beta functions 
\be
\begin{pmatrix}
\dot \kapl
\\
\dot \kapr
\end{pmatrix}
= \tfrac 12 \begin{pmatrix} \kapl \\ \kapr \end{pmatrix}
- d (\kapl^2 + \kapr^2) \begin{pmatrix} \kapl \\ \kapr \end{pmatrix}
\;,
\ee
where $d$ is a scheme-dependent constant. This has the required $O(2)$
symmetry and a ring of fixed points at $\kapl^2 + \kapr^2 = 1/(2d)$.
The minimal models $M_{p,p+1}$ with $p>3$ have non-zero three point
couplings and so the simplest change to these beta functions to
incorporate these is
\be
\begin{pmatrix}
\dot \kapl
\\
\dot \kapr
\end{pmatrix}
= y \begin{pmatrix} \kapl \\ \kapr \end{pmatrix}
- c \begin{pmatrix} \kapl^2 \\ \kapr^2 \end{pmatrix}
- d (\kapl^2 + \kapr^2) \begin{pmatrix} \kapl \\ \kapr \end{pmatrix}
\;.
\ee
This breaks the degeneracy and (for $dy>0$) has six non-trivial fixed
points with the same pattern of flows that we have found, which is
additional evidence in favour of the pattern we propose.

There are also various physical quantities which we would like to
calculate using the TCSA method which for various reasons have proved
intractable, such as the expectation values of $T$ and $\bar T$ (which
would give a better test of a conformal defect than the
examination of the spectrum) or the reflection and tranmission
coefficients defined in \cite{Quella:2006de}.

Finally, it appears likely that a qualitative description of the space
of defects and defect flows can be found from the microscopic RSOS
model description of the minimal models, something we plan to address
in the future.

\vspace{5mm}

\noindent {\bf Acknowledgements} \\[10pt]
We would like to thank 
  Costas Bachas,
  Zolt\'an Bajnok,
  Matthias Gaberdiel,
  G\'abor Tak\'acs,
  G\'abor Zsolt T\'oth and   Paul Fendley 
for interesting discussions on defects and other aspects of this paper.

MK was partially supported by EU grant MRTN-CT-2004-512194 and by
MIUR grant 2007JHLPEZ and thanks the Mathematics department of KCL for
its hospitality.  IR is partially supported by the EPSRC
First Grant EP/E005047/1 and IR and GMTW are partially supported by
the STFC/PPARC rolling grant PP/C507145/1.

\appendix

\sect{Appendix}

\subsection{Defect operators commuting with $D_{(r,1)}$}\label{app:Dr1-orbit}

Let $\delta(s,s')$ be $1/2$ if $s=s'=\tfrac{p+1}2$ and $1$ otherwise, as in the
definition of $F^{s|s'}$ in \eqref{eq:Fss-def}.
To see that $F^{s|s'}$ commutes with the topological defects $D_{(r,1)} 
$ one simply computes
\be\begin{array}{ll}\displaystyle
   D_{(r,1)} F^{s|s'}
   \etb= \sum_{t,u=1}^{p-1} N_{(r,1)(t,1)}^{~~~(u,1)} D_{(u,1)}
   \vvecc{1,s}\ccevv{1,s'} D_{(t,1)}  \cdot \delta(s,s')
\enl
   \etb= \sum_{u=1}^{p-1} D_{(u,1)}
   \vvecc{1,s}\ccevv{1,s'} D_{(u,1)} D_{(r,1)} \cdot \delta(s,s')
   = F^{s|s'} D_{(r,1)} ~.
\eear\ee

To establish the decomposition \eqref{eq:C-fact-rest} of a conformal  
defect that commutes with $D_{(2,1)}$ (and hence with all $D_{(r,1)}$)  
it is enough to show that we can find constants $m_a^{s|s'}$ 
          in \eqref{eq:C-fact-rest2}
such that
\be
   \big( C , \vvecc{a}\ccevv{b} \big)
   = \big( F , \vvecc{a}\ccevv{b} \big)
   \quad \text{for~all}~a,b \in \Ic_p ~.
\labl{eq:C=F-calc1}
Because of the identity $\vvecc{r,s} = \vvecc{p-r,p+1-s}$ it suffices  
to verify this for factorising defects of the form $\vvecc{x,v} 
\ccevv{y,v'}$ with $v \le \tfrac{p+1}2$ and $v' \le \tfrac{p+1}2$.

For the left hand side of \eqref{eq:C=F-calc1} we find
\bea
   \big( C , \vvecc{x,v}\ccevv{y,v'} \big)
= \big( C D_{(y,1)}, \vvecc{x,v}\ccevv{1,v'} \big)
= \big( C , D_{(y,1)} \vvecc{x,v}\ccevv{1,v'} \big)
\enl
= \sum_{a=1}^{p-1} N_{(y,1)(x,1)}^{~~~(a,1)} \big( C ,  \vvecc{a,v} 
\ccevv{1,v'} \big)~.
\eear\labl{eq:C=F-calc2}
To reproduce this result with $\big( F , \vvecc{x,v}\ccevv{y,v'} \big) 
$ we will distinguish four cases,
namely whether $v$ and $v'$ are less or equal to $\tfrac{p+1}2$,  
respectively.
Of course, the three cases where either $v$ or $v'$ are equal to $ 
\tfrac{p+1}2$
can occur only for $p$ odd.
We will abbreviate
\be
   q = \frac{p+1}2 ~.
\ee
For $m_a^{s|s'}$ we make the ansatz
\be
   m_a^{s|s'} = \mu(s,s') \cdot \big( C ,  \vvecc{a,s}\ccevv{1,s'}  
\big) ~,
\labl{eq:mss-def}
where $\mu(s,s') = \tfrac12$ if at least one of $s,s'$ is equal to $q 
$, and $\mu(s,s') = 1$ otherwise.
The $m_a^{s|s'}$ have the property that for all $s,s'$,
\be
   m_a^{q|s'} = m_{p-a}^{q|s'}
   \quad \text{and} \quad
   m_a^{s|q} = m_{p-a}^{s|q} ~.
\ee
The first equality is an immediate consequence of $\vvecc{a,q} =  
\vvecc{p-a,q}$,
while for the second one we need to commute the defect operators  
$D_{(p,1)}$ and $C$
and use $\ccevv{p,q} = \ccevv{1,q}$,
\bea
   m_{p-a}^{s|q}
   = \tfrac12  \big( C ,  D_{(p,1)} \vvecc{a,s}\ccevv{1,q} \big)
   = \tfrac12  \big( C  D_{(p,1)} , \vvecc{a,s}\ccevv{1,q} \big)
\enl
   = \tfrac12  \big( C  , \vvecc{a,s}\ccevv{1,q} D_{(p,1)} \big)
   = \tfrac12  \big( C  , \vvecc{a,s}\ccevv{1,q} \big)
   =  m_{a}^{s|q} ~.
\eear\ee

 From their definition it is easy to check that
the factorising defects $F^{s|s'}$ obey the identities
\be
   F^{p+1-s|p+1-s'} = F^{s|s'}
   \quad , \quad
   D_{(a,1)} F^{p+1-s|s'} = D_{(p-a,1)} F^{s|s'} ~.
\ee
This allows us to restrict the range of the $s$ and $s'$ summation in
\eqref{eq:C-fact-rest2} to $s,s' \le q$. We will now establish the  
equality
\eqref{eq:C=F-calc1} in the four cases.

\medskip\noindent
i) $v,v' < q$: Then
\bea
   \big( F , \vvecc{x,v}\ccevv{y,v'} \big)
=
\sum_{s,s'=1}^{\lfloor q \rfloor} \sum_{a,t=1}^{p-1} m_{a}^{s|s'}  
\delta(s,s')
   \big( D_{(a,1)} D_{(t,1)}
   \vvecc{1,s}\ccevv{1,s'} D_{(t,1)} ,  \vvecc{x,v}\ccevv{y,v'} \big)
\enl
=
\sum_{s,s'=1}^{\lfloor q \rfloor} \sum_{a,t,u=1}^{p-1} m_{a}^{s|s'}
   N_{(a,1)(t,1)}^{~~~(u,1)} \delta(s,s')
   \big( \vvecc{u,s}\ccevv{t,s'} ,  \vvecc{x,v}\ccevv{y,v'} \big)
\enl
=
\sum_{a}^{p-1} \big( C ,  \vvecc{a,v}\ccevv{1,v'} \big)
   N_{(a,1)(y,1)}^{~~~(x,1)} ~,
\eear\ee
where in the last step we used that since $s,v<q$, $\vvecc{u,s} =  
\vvecc{x,v}$ if
and only if $u=x$ and $s=v$, and similarly for $\ccevv{t,s'}$ and $ 
\ccevv{y,v'}$.
Since $v,v'<q$ this implies also $\delta(s,s')=1$.
The above expression is equal to \eqref{eq:C=F-calc2} by
the symmetries of the minimal model fusion rules.

\medskip\noindent
ii) $v<q, v'=q$: Then
\bea
   \big( F , \vvecc{x,v}\ccevv{y,q} \big)
=
\sum_{s,s'=1}^{\lfloor q \rfloor} \sum_{a,t,u=1}^{p-1} m_{a}^{s|s'}
   N_{(a,1)(t,1)}^{~~~(u,1)} \delta(s,s')
   \big( \vvecc{u,s}\ccevv{t,s'} ,  \vvecc{x,v}\ccevv{y,q} \big)
\enl
=
\sum_{a=1}^{p-1} m_{a}^{v|q} \delta(v,q)
   \big( N_{(a,1)(y,1)}^{~~~(x,1)} + N_{(a,1)(p-y,1)}^{~~~(x,1)} \big)
=
\sum_{a=1}^{p-1} m_{a}^{v|q} N_{(a,1)(y,1)}^{~~~(x,1)} +
\sum_{a=1}^{p-1} m_{p-a}^{v|q} N_{(a,1)(y,1)}^{~~~(x,1)}  ~,
\eear\ee
where we used that $N_{(a,1)(p-y,1)}^{~~~(x,1)} = N_{(p-a,1)(y, 
1)}^{~~~(x,1)}$.
Since $m_{p-a}^{v|q} = m_{a}^{v|q}$, the factor of $\tfrac12$ in
\eqref{eq:mss-def} ensures that both sums combine to give  
\eqref{eq:C=F-calc2}.

\medskip\noindent
iii) $v=q, v'<q$: This case works along the same lines.

\medskip\noindent
iv) $v=q, v'=q$: Then
\bea
   \big( F , \vvecc{x,q}\ccevv{y,q} \big)
=
\sum_{s,s'=1}^{\lfloor q \rfloor} \sum_{a,t,u=1}^{p-1} m_{a}^{s|s'}
   N_{(a,1)(t,1)}^{~~~(u,1)} \delta(s,s')
   \big( \vvecc{u,s}\ccevv{t,s'} ,  \vvecc{x,q}\ccevv{y,q} \big)
\enl
=
\sum_{a=1}^{p-1} m_{a}^{q|q} \cdot \tfrac12 \cdot
   \big(
    N_{(a,1)(y,1)}^{~~~(x,1)} + N_{(a,1)(p-y,1)}^{~~~(x,1)} +
    N_{(a,1)(y,1)}^{~~~(p-x,1)} + N_{(a,1)(p-y,1)}^{~~~(p-x,1)}
   \big)
\enl
=
\sum_{a=1}^{p-1} m_{a}^{q|q}
   \big(
    N_{(a,1)(y,1)}^{~~~(x,1)} + N_{(a,1)(p-y,1)}^{~~~(x,1)} 
   \big)
\eear\ee
where we used that $N_{(a,1)(p-y,1)}^{~~~(p-x,1)} = N_{(a,1)(y, 
1)}^{~~~(x,1)}$.
The rest of the argument is as in ii).

\newpage
\subsection{The renormalisation group and finite-size scaling relations in TCSA}
\label{app:TCSA}

In this section we derive the form of the finite-size scaling flow in
TCSA given in equations \eqref{eq:fss1}--\eqref{eq:fss3}.
We start from the three assumptions in section \ref{sec:fss}:
if $\kapp{i}(N,t)$ are the TCSA coupling constants along a finite-size
scaling flow parametrised by $t$ at cut-off $N$, then we assume
\begin{eqnarray}
 \kapp{i}(\infty,t) 
&=& e^{yt}\;\kappp{i}{0}
\;,
\label{eq:fss4}
\\
 - N \frac{\partial \kapp{i}}{\partial N}(N,t)
 &=& \bt{i}(\kappa(N,t);N)
\;,
\label{eq:fss5}
\\
  \bt{i}(\kappa;N) &=& N^y \gamm{i}(\kappa N^{-y}) 
\;.
\label{eq:fss6}
\end{eqnarray}
Firstly we remove the expected $t$ dependence from $\kapp{i}(N e^t,t)$
and define
\be
 \sigma^i(N,t) = e^{-yt} \kapp{i}(N e^t,t)
\;.
\labl{eq:sigma0}
This satisfies the differential equation
\be
 - N \frac{\partial}{\partial N}  \sigma^i(N,t) 
= N^y \gamm{i}(N^{-y}\sigma)
\;,
\labl{eq:sigma1}
with initial conditions
\be
 \updown{\sigma}{i}(\infty,t) = \kappp{i}{0}
\;.
\labl{eq:sigma2}
Since both the initial conditions and the differential equation are
independent of $t$, the solution is also independent of $t$ and we
find that
\be
  \updown{\sigma}{i}(N,t) = \updown{f}{i}(\kappp{j}{0} ;N)
\;.
\labl{eq:sigma3}
Substituting this into \eqref{eq:sigma0} we find our first result,
\be
 \kapp{i}(N e^t,t) = e^{yt}  \updown{f}{i}(\kappp{j}{0};N) = e^{yt} \kapp{i}(N,0)
\;.
\ee
Rescaling $N$ in this equation gives the second result
\be
 \kapp{i}(N ,t) = e^{yt}  \kapp{i}(N e^{-t},0)
\;.
\ee
Finally, differentiating this with respect to $t$ at fixed level $N_0$
gives
\begin{eqnarray}
    \frac{\partial}{\partial t}\kapp{i}(N_0,t) 
&=& y  e^{yt}  \kapp{i}(N_0 e^{-t},0) -  e^{yt}N_0
e^{-t}\frac{\partial\kapp{i}}{\partial N} (N_0 e^{-t},0)
\nonumber\\
&=& y \kapp{i}(N_0,t) + e^{yt}\bt{i}(\kapp{i}(N_0e^{-t},0);N_0e^{-t})
\nonumber\\
&=& y \kapp{i}(N_0,t) + N_0^y \gamm{i}(N_0^{-y} e^{yt}\kapp{i}(N_0e^{-t},0))
\nonumber\\
&=& y \kapp{i}(N_0,t) + \bt{i}(e^{yt}\kapp{i}(N_0e^{-t},0);N_0)
\nonumber\\
&=& y \kapp{i}(N_0,t) + \bt{i}(\kapp{i}(N_0,t);N_0)
\nonumber\\
&=& \bb{i}( \kappa(N_0,t) ; N_0)
\;.
\end{eqnarray}

\subsection{Some details of the TCSA algorithm}
\label{sec:TCSA_app}

In order to calculate the matrix elements of the perturbations one needs to
know the functional form of the 3pt function of primary fields on the upper
half plane. If the perturbing operator is chiral it is simply
\be
\3pt{A}{\phi_h(z)}{B}=C\,z^{h_A-h-h_B}\,.
\labl{eq:3ptchiral}
If the perturbation is non-chiral then the correlation function is more
complicated, but if the states $A$ or $B$ belong to the representation $(1,2)$
or $(2,1)$ the null vector equations can be used to deduce the form of the correlator.

The equation in the model $M(p,q)$ for $B$ is
\be
\3pt{A}{\phi(z,\bar z)(L_{-1}^2-tL_{-2})}{B}=0
\ee
where $t=p/q$ if $B=\phi_{1,2}$ and $t=q/p$ if $B=\phi_{2,1}$. Commuting the
$L_n$'s past $\phi$, changing to polar coordinates $(r,\theta)$ and using
\be
(h_A-h_B)\3pt{A}{\phi_{h,\bar h}(r,\theta)}{B}=\3pt{A}{[L_0,\phi_{h,\bar
      h}(r,\theta)]}{B}= \left(h+\bar{h}+r\frac\partial{\partial r}\right)\3pt{A}{\phi(r,\theta)}{B}
\ee
one arrives at the second order differential equation
\begin{multline}
\left[\sin^2\theta\,\,\partial_\theta^2+(1+\Delta-t)\sin2\theta\,\partial_\theta+\right.\\
\left.\Delta(\Delta+1)\cos^2\theta-\Delta\sin^2\theta-t(\Delta\cos2\theta+h\,
e^{-2i\theta}+\bar h\,e^{2i\theta})\right]\3pt{A}{\phi_{h,\bar h}(r,\theta)}{B}=0
\label{eq:nulleq1}
\end{multline}
where $\Delta=h_A-h-\bar h-h_B$.

The null vector equation for $A$ reads
\be
\3pt{(L_{-1}^2-tL_{-2})A}{\phi(z,\bar z)}{B}=0\,.
\ee
With the same steps this can be converted to the equation
\begin{multline}
\Big[\sin^2\theta\,\,\partial_\theta^2+2\sin\theta\left((1+2h+2\bar
h-t-\Delta)\cos\theta+2i(h-\bar h)\sin\theta\right)\,\partial_\theta +\\
e^{2i\theta}\left(4h^2-h(3t+2\Delta-2)\right)+e^{-2i\theta}\left(4\bar
h^2-\bar h(3t+2\Delta-2)\right)+\\
\frac12\Delta(\Delta+2t-2)\cos(2\theta)+\frac12(4h-\Delta)(4\bar
h-\Delta)\Big]\3pt{A}{\phi_{h,\bar h}(r,\theta)}{B}=0\,.
\label{eq:nulleq2}
\end{multline}

If it is not true that $h_A=h_B$ {\it and} $h=\bar h$ at the same time
the equations \erf{eq:nulleq1} and \erf{eq:nulleq2} can be combined to
obtain a first order equation. In case $h_A=h_B$ and $h\neq\bar h$ one
gets
\be
\left[\partial_\theta+(1+h+\bar h-2t)\cot\theta+i(h-\bar
  h)\right]\3pt{A}{\phi_{h,\bar h}(r,\theta)}{B}=0
\ee
having the solution
\begin{align}
\3pt{A}{\phi_{h, \bar h}(r,\theta)}{B}&=C\,r^{-h-\bar h}e^{-i(h-\bar
  h)\theta}\sin\theta^{-1+2t-h-\bar h}=\\
&=\tilde Cz^{(1-2t-h+\bar h)/2}\bar z^{(1-2t+h-\bar h)/2}(z-\bar z)^{-1+2t-h-\bar h}\,.
\label{eq:3ptnonchir}
\end{align}

The matrix elements of the perturbation in the basis
\erf{eq:basis} can be calculated using the relations
\begin{align}
\3pt{L_{-n}A}{\phi(z,\bar z)}{B}&=
\sum_{k=-1}^n\binom{n+1}{k+1}\left(z^{n-k}\3pt{A}{L_k\phi(z,\bar z)}{B}+\bar
z^{n-k}\3pt{A}{\bar L_k\phi(z,\bar z)}{B}\right)+\nonumber\\
&\3pt{A}{\phi(z,\bar z)}{L_nB}\,,\\
\3pt{A}{\phi(z,\bar z)}{L_{-n}B}&=
-\sum_{k=-1}^\infty\binom{-n+1}{k+1}\left(z^{-n-k}\3pt{A}{L_k\phi(z,\bar
  z)}{B}+\bar z^{-n-k}\3pt{A}{\bar L_k\phi(z,\bar z)}{B}\right)+\nonumber\\
&\3pt{L_nA}{\phi(z,\bar z)}{B}\,.
\end{align}
For $\phi$ left chiral the terms containing $\bar z$ are not
present. Applying these relations iteratively every matrix element becomes a
sum of correlators of the form 
\be
\3pt{A}{L_{-1}^k\bar L_{-1}^l\phi(z,\bar z)}{B} \qquad\qquad A,B\text{ primary.}
\ee
The action of $L_{-1}$'s translates to derivatives of \erf{eq:3ptchiral} or
\erf{eq:3ptnonchir} with respect to $z$ and $\bar z$.

\subsection{Position invariance of the spectrum for a chirally perturbed defect}
\label{app:TCSA-posn-indep}

Let $H(\theta)$ be the Hamiltonian in \eqref{eq:pert-Ham} for the perturbation
by a single chiral field $\phi_{h,0}$,
\be
H(\theta)=H_0 + H_I(\theta) 
~~\text{where}~~
H_0 = \frac\pi{R}\Big(L_0-\frac{c}{24} \Big)
~,~~
H_I(\theta) = 
  \laml
  \left(\frac{\pi}{R}\right)^{\!h}
  e^{i h \theta} \,
  \phi_{h,0}(e^{i\theta}) \,.
\ee
We will show that 
\be
  H(\theta) = e^{i \theta L_0} H(0) e^{-i \theta L_0}
\labl{eq:H(th)-simtrans}
by verifying it on matrix elements. 
Clearly, $e^{i \theta L_0} H_0 e^{-i \theta L_0} = H_0$, so that it is enough to show
$H_I(\theta) = e^{i \theta L_0} H_I(0) e^{-i \theta L_0}$. On the one hand, from 
\eqref{eq:3pt-coord-dep} we have, in the notation of that section,
\be
\3pt{v_i}{H_I(\theta)}{v_j}= \laml\,
  \left(\frac{\pi}{R}\right)^{h}\,
  e^{i h \theta} \, C\,e^{i\theta(\Delta_i-h-\Delta_j)}\ , 
\ee
where the constant $C$ is given by $C = \3pt{v_i}{\phi_h(1)}{v_j}$. On the other hand,
\be
  \3pt{v_i}{e^{i \theta L_0} H_I(0) e^{-i \theta L_0} }{v_j}
  = e^{i \theta (\Delta_i - \Delta_j)} \3pt{v_i}{ H_I(0) }{v_j} 
  = e^{i \theta (\Delta_i - \Delta_j)} 
    \laml
  \left(\frac{\pi}{R}\right)^{\!h}
  \3pt{v_i}{ \phi_{h,0}(1) }{v_j} 
\ee
These two expressions coincide.
We see that $H(\theta)$ and $H(0)$ are related by a similarity
transformation, and hence have the same spectrum. For the application
to TCSA it is important to note that the operator $e^{i \theta L_0}$
commutes with the projection $P_N$ to the truncated Hilbert space, so
that also 
\be
  P_N H(\theta) P_N = e^{i \theta L_0} P_N H(0) P_N e^{-i \theta L_0} ~.
\ee
Thus even in TCSA, the truncated Hamiltonians for different positions
$\theta$ of the chirally perturbed defect on the strip have {\em
  exactly} the same spectrum. For a perturbation by an anti-chiral
field $\phi_{0,h}$, the argument is the same, except that the
similarity transformation in this case is $H(\theta) = e^{-i \theta
  L_0} H(0) e^{i \theta L_0}$.

\newcommand\arxiv[2]      {\href{http://arXiv.org/abs/#1}{#2}}
\newcommand\doi[2]        {\href{http://dx.doi.org/#1}{#2}}
\newcommand\httpurl[2]    {\href{http://#1}{#2}}

\end{document}